\documentclass[aps,amsmath,prb,a4paper,floatfix,twocolumn]{revtex4}
\usepackage{graphicx}
\usepackage{subfigure}
\usepackage{color}
\usepackage{amsmath}
\usepackage{amsfonts}
\usepackage{amssymb}
\newcommand{\beq}{\begin{equation}}
\newcommand{\eneq}{\end{equation}}
\newcommand{\bea}{\begin{eqnarray}}
\newcommand{\enea}{\end{eqnarray}}

\newcommand{\nubf}{{\bf \nu}_}
\newcommand{\jbf}{{\bf J}}
\newcommand{\ebf}{{\bf {\cal{E}}}}
\newcommand{\tbf}{(-{\bf {\nabla}} T)}
\newcommand{\entalp}{\epsilon + {\cal{P}}}

\newcommand{\tx}{{\tilde{x}}}

\begin{document}
\title{Thermal transport driven by charge imbalance in graphene  in magnetic field, close to the charge neutrality point  at low temperature: Non local resistance }
\author{ A.Tagliacozzo$^{1,4}$,G.Campagnano$^{1}$, D.Giuliano$^{2,4}$,P.Lucignano$^{4}$, B.Jouault$^{3}$}
\affiliation{$^{1}$  Dip. di Fisica, Universit\`{a} di Napoli Federico II, Via Cintia, I-80126 Napoli, Italy}
\affiliation{$^{2}$  Dipartimento di Fisica, Universit\`a della Calabria Arcavacata di Rende I-87036, Cosenza, Italy}
\affiliation{$^{3}$   Universit\'e Montpellier-CNRS, Laboratoire Charles Coulomb UMR 5221, F-34095, Montpellier, France }
\affiliation{$^{4}$ CNR-SPIN, Monte S.Angelo via Cinthia, I-80126 Napoli, Italy}
%\affiliation{$^{5}$  INFN, Gruppo collegato di Cosenza, Arcavacata di Rende I-87036, Cosenza, Italy}
%\affiliation{$^{5}$  Department of Mathematics, City, University of London, EC1V 0HB London, United Kingdom}
%\affiliation{$^{6}$  CNR-SPIN, Monte S. Angelo-Via Cintia, I-80126, Napoli,  Italy }

\begin{abstract}
Graphene grown epitaxially on SiC, close to the charge neutrality point  (CNP), in an orthogonal magnetic field 
shows an ambipolar behavior  of  the transverse resistance  accompanied by a puzzling  
longitudinal magnetoresistance. When injecting a transverse current at one end of the Hall bar, a sizeable  non 
local  transverse magnetoresistance is measured at low temperature. While Zeeman spin effect seems not to be able to  
justify these phenomena,  some dissipation involving edge states at the boundaries could explain the order of magnitude of the 
non local  transverse magnetoresistance, but not  the  asymmetry when  the orientation of the orthogonal magnetic field is reversed.
As a possible contribution to the explanation of the measured non local magnetoresistance  which is odd in the magnetic field,  we derive  a hydrodynamic 
approach to transport in this system, which involves particle and hole  Dirac  carriers, in the form of  charge and energy currents. 
We find that thermal diffusion can take place on a large distance  scale, thanks to long recombination times, provided a  non insulating 
bulk of the Hall bar is assumed, as recent models  seem to suggest in order to explain  the appearance of the longitudinal resistance. In 
presence of the local source,  some leakage of carriers from the edges generates an imbalance of carriers of opposite sign, 
which are  separated in space by the magnetic field and  diffuse along the Hall bar generating a non local transverse voltage. 

\vspace{0.5cm}

%  10  April 2017,  13  June 2018, 10  July 2018, 24 January  2019
\end{abstract}

\maketitle

\section{Introduction}
\setcounter{equation}{0}
   Taming quantum  transport   at the edges  of high mobility graphene Hall bars  provides 
   control of the quantization of the fractional Hall effect  and is the prerequisite for the 
   implementation of  non-Abelian braiding statistics of excitations, which has been proposed as a 
   tool for alternative quantum information processing\cite{Nayak2008}. 
 In a two-dimensional electron gas, Quantum Hall Effect is a nowadays 
 paradigmatic example of a ``bulk'' incompressible insulating phase, with 
compressible chiral edge states, carrying a charge either integer (Integer Quantum Hall Effect) 
\cite{edgeinteger},  or fractional (Fractional Quantum Hall Effect) \cite{edgefractional}. 
In addition to chiral current-carrying states, at so-called ``non-Laughlin fillings'', 
such as $\nu = 2/3$, additional counterpropagating 
neutral boundary states have been predicted \cite{23_1,Deviatov2011,Granger2009}, which can in principle be detected 
by e.g looking at the extra charge noise they generate, once excited \cite{rosenow,cheianov},
by means of momentum-resolved tunneling \cite{stern},  or measuring 
current-current correlations in a pertinent generalization of the 
quantum point contact scatterer between edge states proposed in \cite{campa_1}. 
Evidence for the  neutral edge modes has recently been provided in a simultaneous 
measurement of both the chemical potential and temperature at a quantum Hall edge 
heated by means of a quantum point contact. As a result, it has been found that, while 
the charge is exclusively transported downstream, when the edge is expected to 
have additional structures such as the neutral branch of counterpropagating modes 
at $\nu = 2/3$, heat can be transported upstream. In addition, 
an unexpected bulk contribution to heat transport was also found  at Integer 
Quantum Hall fillings, in particular at $\nu =1$ \cite{Venkatachalam2012}.  
%Similar effects have been found in   two dimensional electron gas of metal-oxide-semiconductor heterostructures\cite{Deviatov2011,Granger2009}.  
In graphene Hall bars in an external magnetic field, the observation of large nonlocal resistances near the Dirac point   adds fuel to the mystery 
about the role of the external confining potential that defines the edge. 

 Clean graphene at the  Charge Neutrality Point ( CNP)  attracts a lot of attention because  it can provide an example of a 
 strongly-interacting quasi-relativistic electron- hole plasma, known as a Dirac fluid\cite{Sheehy2007}. Coulomb interaction is 
 controlled by the dimensionless parameter  $ \alpha _f$ and the only relevant energy scale is temperature $T$ which determines the  
 inelastic  scattering rate $\tau_{c}$,
 \bea
  \tau_{c}^{-1}  \sim \alpha_f^2  \frac{k_BT}{\hbar}    \:\:\:\:  with  \:\:\:\: \alpha_f =   \frac{ e^2}{\epsilon _r \hbar v_F} \sim \frac{2}{\epsilon _r} \sim 1 .
  \label{taue}
\enea
Eq.(\ref{taue}) is a hallmark of many quantum-critical systems \cite{Sheehy2007,Fritz2008,Mueller2008a}. While in a Fermi liquid 
the heat current is  strictly related to the mass current, close to the CNP  the thermal conductivity is expected to  be enhanced, because 
particles and holes move both in  the direction of the thermal gradient with a strongly reduced  relaxation rate, due to the linearity of the  energy 
dispersion relation \cite{Foster2009}.  
Backscattering of Dirac fermions is also suppressed if intervalley scattering, which comes into play only in presence of strong disorder,  
is neglected\cite{Ando1998}.  Violations of the Wiedemann Franz law\cite{Ashcroft} have been reported, as well,  as a consequence of the strong  Coulomb 
interactions between thermally excited charge carriers at the CNP\cite{Crossno2016}.
 Noticeably, a strong enhancement of the Lorentz number by a factor of about 2.5 its 
Wiedemann-Franz value has also been derived in a ballistic bilayer graphene in the presence
of trigonal warping term and  with the electrochemical potential  close to the Lifshitz energy, as 
a consequence of the van Hove singularities in the single particle density of states \cite{susz}.

%  While at thermal equilibrium  here heat flow can take place even in the absence of matter convection, a consequence of particle-hole creation and annihilation.  
 
In a Dirac fluid,  the  violations of the Wiedemann Franz law have been reported  in a temperature window  
$\frac{ \hbar}{k_B \tau_{imp} } < T< \frac{ \hbar}{k_B \tau_{ph} } $, where 
$1/  \tau_{ph} $ refers to the relaxation mechanism  introduced by inelastic phonon scattering at higher temperatures. 
The reason for this is that   charged impurities in the substrate  generate  a local finite carrier concentration 
 (particle or hole puddles \cite{DasSarma2011}), which can give rise to a position dependent chemical potential  $\mu (r)$, so that  
 both electric and thermal  transport may be dominated by  an elastic  scattering rate, 
\[  \tau_{imp}^{-1}  \sim \left ( \frac{Ze^2}{\epsilon _r}\right ) ^2 \frac{ \rho_{imp}}{ \hbar \:  max[ \mu, k_BT]}.   \] $ \tau_{imp}^{-1} $
 is naturally proportional to the impurity density  and is responsible for restoring a Fermi liquid-like behavior.  At larger doping, when 
the chemical potential   exceeds $k_BT$, the inelastic-scattering rate tends to the familiar
Fermi-liquid form  $\sim T^2/\mu $   if the interactions are screened. 
Favourable  experimental  bath temperatures to  monitor the Dirac fluid properties in  boron nitride (hBN) encapsulated  
graphene has been found to be\cite{Crossno2016} $T> 40 \:^oK$. 

Regarding charge transport,  the longitudinal conductance  is known to be finite  at the CNP, in the absence of any disorder:  
$ \sigma _Q = 4 e^2 /\pi h  $ \cite{Katsnelson2006,Tworzydlo2006}. Indeed, the Zitterbewegung, with the creation of virtual zero total momentum electron hole pairs, 
appears to be responsible for the evanescent zero energy Dirac modes,  which provide the finite conductivity.\cite{Damle1997,katsnelson2012graphene}

\begin{figure}
\begin{center}
\includegraphics[width= 0.9 \linewidth]{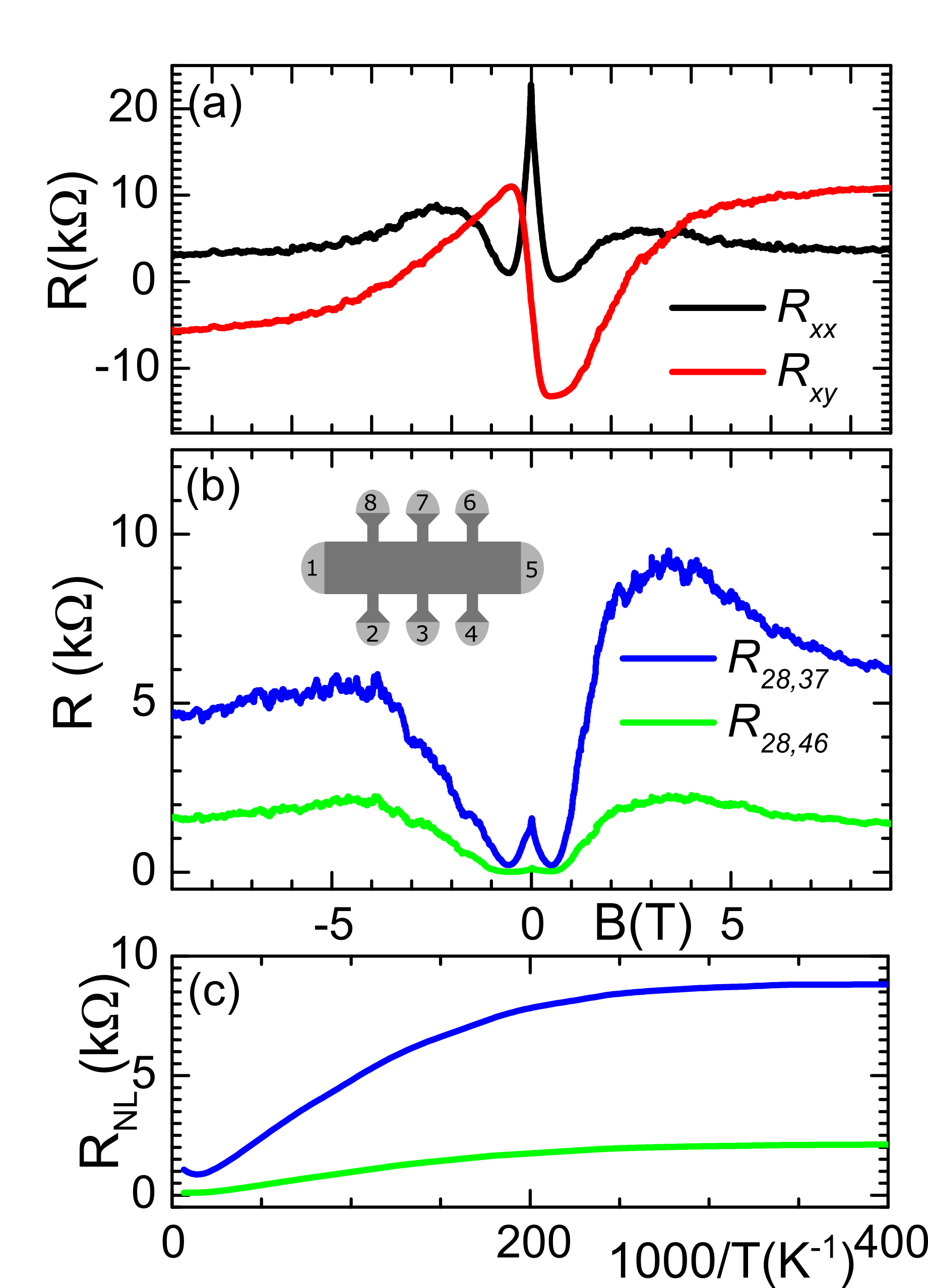}
\caption{(a) Longitudinal (black line) and transverse (red line) magnetoresitances $R_{xx}$ and $R_{xy}$ as measured on a 
graphene Hall bar close to the CNP at $T=$ 1.7 K and $I=$ 10 nA. (b) Non local magnetoresistances $R_{28,37}$ (close to the 
current source, blue line) and $R_{28,46}$ (far from the current source, green line). The inset is a sketch of the Hall bar 
which labels the contacts. (c) Temperature dependence of the nonlocal resistances $R_{28,37}$ and $R_{28,46}$ at a fixed magnetic field $B=3$ T.}
\label{tef23}
\end{center}
\end{figure}

\begin{figure}
\begin{center}
\includegraphics[width= 0.7 \linewidth]{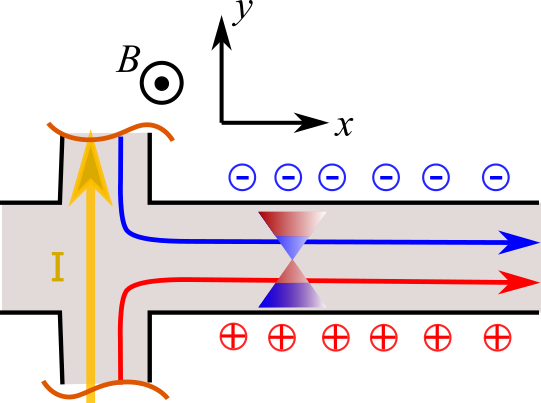}
\caption{ Sketch of the nonlocal experiment. A charge flow is applied across the Hall bar on its left part (yellow colour). 
The magnetic field induces space separation of the charges of opposite sign, preferentially leaking from the edges  (blue and read circles) 
in the bulk of the Hall bar. Thermal diffusion along the Hall  bar drives the charge imbalance of the chemical potential between 
electrons (blue line)  and holes (red line)  away from the source, which relaxes poorly  over microscopic distances.  The thermal 
and Lorentz force acting on the imbalance flow are compensated by a transverse electric field which can be detected by nonlocal probes.
Flipping the direction of the magnetic field requires exchanging the read and blue circles and lines with each other.}
\label{scheme3}
\end{center}
\end{figure}

In this paper, we are mainly interested in giant nonlocal voltages
appearing in graphene in the QHE regime close to the CNP at low temperature and in
the presence of a magnetic field.
These nonlocal voltages have been observed by various authors~\cite{Abanin2011, Gorbachev2014, Renard2014, Ribeiro2017},
including some of us.\cite{Nachawaty2018}, though under operating conditions rather far from  the two limits (Dirac and Fermi liquid behavior)  
mentioned above.  However,  here we claim   that one cannot explain the 
unusual electric properties found  close to  the CNP without accounting for 
thermal diffusion, even at such low temperatures.  

For the sake of clarity, we now present some typical experimental results,
complementary of those shown in Ref.[\onlinecite{Nachawaty2018}].
Figure~\ref{tef23}a shows the longitudinal and transverse 
magnetoresistances observed  in a graphene Hall bar epitaxially grown onto SiC.
The magnetic field is orthogonal to the sample plane,
the temperature is $T=$ 1.7 K and
the Fermi energy is tuned close the CNP (Hall concentration $p \simeq 10^{10}$ cm$^{-2}$, mobility $\simeq 10,000$~cm$^2$/Vs).
The graphene Hall bar is macroscopic and has a length
$\ell \simeq $ 400 $\mu$m and a width $w \simeq$ 100 $\mu$m.
The peak in  the longitudinal resistance $R_{xx}$ at $B=0$ is due to weak or strong localization and its discussion 
is out of the scope of this paper.
When the  magnetic field is turned on, this peak disappears, while another peak can be resolved
around $B=\pm 2.5$ T. At approximately the  same value of $B$, the transverse resistance
has an ambipolar behavior and goes to zero.
We conclude  that for some reasons (intrinsic to graphene on SiC),
the Fermi energy smoothly increases   with $B$ and crosses the CNP around
$B= 2.5$T.
%The $R_{xx}$ peak around $B=\pm 3$ T corresponds to the diffusive longitudinal peak observed at the CNP in Ref.~\onlinecite{Abanin2011}.
%
More remarkably, Figure~\ref{tef23}b shows the nonlocal resistance measured  across the bar at two different places, 
when the Hall bar is  current biased at one of its extremities, along its width (transverse direction), while it  
is kept as an open circuit along its length (longitudinal direction).  
Here we define the resistance  $R_{ij,kl}= V_{kl}/I_{ij}$,
where $V_{kl}$ is the voltage drop between contacts $k$ and $l$,
and  $I_{ij}$ is  the current biased between contacts $i$ and $j$ (see inset in  Fig.\ref{tef23}b).
%The current is now applied across the Hall bar.
%
Remarkably, the Ohmic resistance $R_{Ohmic}$ decreases with the distance from the 
applied current   much more slowly than one would as one would naively expect from 
the classical spreading  of the charge flow inside the Hall bar, which would 
yield 
\beq
R_{Ohmic} = \frac{4}{\pi}  \: \rho _{xx} \: e^{-\pi  D/w}
\label{eq:ohmic}
\eneq
where $\rho_{xx}$ is the resistivity and  $D$ is the distance separating the current injection 
point from the voltage detection  (see Fig.\ref{tef23}b).  
The measured nonlocal resistances observed in Fig.~\ref{tef23}b are
much larger than the ones predicted by Eq.~\ref{eq:ohmic}:
at $B=3$~T, %$R_{xx} \simeq$ 5 k$\Omega$, 
Eq.~\ref{eq:ohmic} predicts
$R_{28,37} \simeq 300$ $\Omega$ and 
$R_{28,46} \simeq 10$ $\Omega$ 
while experimentally one finds  
$R_{28,37} \simeq 10$ k$\Omega$ and 
$R_{28,46} \simeq 2$ k$\Omega$.
Therefore, some other explanation must be found.

If the channel length matches the spin diffusion length in graphene,
the nonlocal configuration makes it possible to detect   spin-related signals in the
spin Hall regime.
However, the distance between probes 2 and 4 for the Hall bar presented in Fig.~\ref{tef23} is 200 $\mu$m, which  
is much larger than the usual spin relaxation lengths reported in graphene.
%
%The nonlocal signals are so large that they are hardly explained by spin Hall effect. also give unrealistically long spin relaxation lengths~\cite{Ribeiro2017, Nachawaty2018}. Zeeman spin Hall effect was proposed as an alternative explanation.However, 
%
Nevertheless, the nonlocality was first attributed to the existence of a Zeeman spin Hall effect (ZSHE), 
where the combination of both Lorentz force and Zeeman spin splitting gives rise to a spin imbalance 
which propagates along the Hall bar.\cite{Abanin2011a}
% relation between Zeeman splitting and the two holes and electrons particles.
%
Later on, additional experimental works demonstrated that
a large part of the nonlocal voltage is insensitive to the magnetic field parallel to the graphene plane - 
suggesting the predominance of orbital effects over ZSHE.

Thermal effects can also give rise to nonlocal voltages. At
the excitation point, heat flow is induced by Ettingshausen
and Joule effects, perpendicular to the charge flow.
This heat flow diffuses and  produces a nonlocal voltage 
far from the charge current, 
via  Nernst effect.\cite{Renard2014,Gopinadhan2015}
However, the observed nonlocal resistances yield
Nernst coefficients which seem to be unrealistically large 
(700 $\mu$V/K in Ref.[\onlinecite{Renard2014}],
more than 10 mV/K in Ref.[\onlinecite{Nachawaty2018}]).

In two dimensional electron gases, 
nonlocal resistances in the QHE regime 
were first observed by McEuen {\it et al.}~\cite{McEuen1990}
and unambiguously attributed to a coupling of 
the edge and bulk conducting pathways.
Later, the same model was used to explain the appearance of nonlocal voltages
in other 2D topological insulators\cite{Gusev2012}.
This model was also extended to graphene near the CNP,
to explain dissipative QHE~\cite{Abanin2011}
and nonlocal voltages~\cite{Ribeiro2017, Nachawaty2018}.
In Ref.[\onlinecite{Abanin2011}],
%the longitudinal resistivity $\rho_{xx}$ has been found in the QHE regime close to CNP\cite{Abanin2007}. Indeed, 
a plateau for the Hall conduction at filling factor $\nu =0 $ has been found, 
accompanied by a peak $\rho_{xx}>0 $ centered at $\nu =0 $. It was argued that this peak is originated by
Zeeman spin splitting of the former  $\nu =0 $ Landau level  centered at fillings  $\nu = \pm 1/2 $. 
The resistance  $\rho_{xx} > 0$ accompanying the unconventional $\nu =0 $ plateaus could be  the result
of interference of edge states located at opposite  boundaries via the  bulk sandwiched  in between,  
having  diffusive conductance. A simple semiclassical description of the interference provides a convincing model 
(henceforth named ``Abanin {\it et al.} model")\cite{Abanin2007}. 

% The same approach has been assumed to be valid  for  the HgCdTe QW to interpret the measured conductances. 
%
The same approach can be used to explain nonlocality. Indeed,
the  peak in $R_{xx}$ around $B=\pm 2.5$ T in Fig.~\ref{tef23}b corresponds to the diffusive
longitudinal peak observed at the CNP in Ref.[\onlinecite{Abanin2011}].
In Ref.[\onlinecite{Nachawaty2018}], a numerical model, again based on the coupling of both edge and 
bulk pathways, was used. Quantitatively, an appropriate choice of the conductivities of the  carriers 
(particles and holes  close to the CNP) could explain most of the non-local resistances.

%Zeeman spin Hall effects  and thermal effects have been excluded\cite{Gusev2012,Nachawaty2017}. Zeeman Spin Hall effect, seem to be ruled out\cite{Renard2014} mostly because the NLR appears to be independent of the magnitude of the magnetic field component, parallel to the graphene flake. The  thermal activation was also considered to be unlikely , because an expected  Nernst coefficient $e_N$ extracted from a Fermi liquid approach is not expected to be larger than $ 1 mV/\:^oK$, which is not enough to invoke thermal effects. Besides, if edge states are responsible for the non locality and circulate around the sample, then they should impose all contacts  to have the same null potential.  
%
%A phenomenological interpretation  has been given, by arguing that counter-propagating edge states are  partly backscattered by the contacts and the bulk of the Hall bar. Quantitatively, an appropriate choice of the conductivities of the  carriers (particles and holes  close to the CNP)  can explain most of these non-local resistances.
%
However, remarkably, the nonlocal resistances are also asymmetric under flipping  $B$\cite{Ribeiro2017}, and 
the closer the measuring contacts are  to the applied bias, the larger is the asymmetry. 
This can be readily observed in Fig.~\ref{tef23}b.
The model for the NLR in terms of dissipative scattering of the edge states \cite{Nachawaty2018} accounts for 
the magnitude of the effect but  does not  catch the origin of the asymmetry when  the $B$ field  is reversed. 
In Ref. [\onlinecite{Ribeiro2017}],
an explanation for the asymmetry has been proposed, assuming that grain boundaries in the graphene sheet have different transmission either for spin up or down. 
This explanation is valid for the experiments of Ref.[\onlinecite{Ribeiro2017}],
which were performed on polycristalline graphene.   However, the $+B/-B$ asymmetry is also clearly visible in Fig.~\ref{tef23}b
 and in Ref.[\onlinecite{Nachawaty2018}], where graphene has been obtained by epitaxy on SiC and does not contain grain boundary. 
 Also, edge dislocations could play a role, as well \cite{parente}. 
	
The purpose of our work is to propose another and more generic approach to justify further contributions to the nonlocal voltages and their asymmetry.
In line with the approach proposed by Ref.[\onlinecite{Abanin2007}],  we consider 
transport in the interior (henceforth named ``bulk'') of the graphene Hall bar away from the edges by reconsidering thermal effects.
 Close to particle-hole symmetry point,  both particles  ($e^-$) and holes  ($h^+$) contribute to transport. 
Here we explore the possibility that the non-equilibrium conditions induced by the  electric field ${\cal{E}}$  
applied at the one end of the Hall bar and Lorenz force  could give rise to local  charge imbalance  between particles and holes  
which could have long relaxation times \cite{Foster2009,Rana2007} at low temperature and could diffuse under the action of 
Joule heating and the thermal gradient generated by the injected current (Ettingshausen effect),
far away from the source. The contribution to the non-local voltage produced in this way depends on the orientation 
of the magnetic field because particles and holes exchange their role when the field is flipped. 
This  could be the origin of the anisotropy when the $B$ field is reversed.  If this interpretation holds, 
the magnitude of the anisotropy is a measure  of the relative weight between the  contribution to dissipation 
in the transport at  the edges  and the thermal diffusion  of the charge imbalance in the bulk. 
 
 The bulk carriers may be created by leakage from the edge states into the bulk as assumed in the Abanin {\it et al.} model. 
 Charged impurities in the substrate which may have a preferred charge and  weak disorder  could also  locally enhance the charge imbalance\cite{Hering2015}.
Or intrinsically,  by adding next-nearest-neighbor hopping terms  in the tight-binding calculation for graphene bands, the Fermi velocity  
at the Dirac cone can  be different between particles and holes. Elastic scattering is ineffective in reducing the charge imbalance. On the other hand,  
thermal inelastic scattering may produce particle hole pairs with different relaxation times, $  \tau_{c} \to   \tau_{ee},  \tau_{hh},  \tau_{eh}$, 
so that the charge imbalance does not relax.  e-ph scattering, as a source of imbalance relaxation is ruled out  at low temperatures.

 Close to the quantum critical point ( $\mu , T \sim 0 $), there is an emergent  relativistic invariance of the interacting 
 electronic Hamiltonian  for the clean sample. In this regime,  a relativistic hydrodynamical approach is expected to apply   as long as   
 $\omega \tau_{c}   << 1 $ (where $ \tau_{c} $ is the inverse cyclotron frequency).

The paper is organised as follows:

In Section \ref{sec2} we first review standard results about  the 
linear response to   external perturbations 
including electric and thermal gradient.  Therefore, we extend those results to systems with two  types of charge carriers,  
$ e^- $ and  $ h^+$.   The Boltzmann transport theory, which applies well to  the  Fermi Liquid regime $\mu >> T $, 
cannot be  safely applied here, because $\mu \sim T $. 
Nervertheless, assuming the perturbation to be small, the  relativistic Dirac electron picture and  the  Lorenz invariant  relativistic equations
of motion  are still valid in the linear approximation, because the constraints posed by 
the conservation laws on the  energy-momentum tensor including dissipative processes 
( viscosity and thermal conduction)\cite{Landau}. 
In the limit in which the disordered impurity distribution  is smooth on the scale of the lattice constant, 
one can assume that the charge imbalance can be described by two different  chemical potentials $\mu_e , \mu_h $, respectively 
for $ e^- $ and  $ h^+$.    At equilibrium, one has $\mu_e = -\mu_h = \mu $ and, accordingly,  in this case 
one can  define the electro-chemical potential $ \mu= (\mu_e - \mu_h)/2 $ and 
the  imbalance chemical potential $ \mu_I = (\mu_e + \mu_h)/2 $. 
  
Section \ref{sec3} reviews the  model of Ref.[\onlinecite{Abanin2007}] for the leakage of edge current into the bulk of the Hall bar. 
In particular, while the original version of the model focuses onto  the linear response to an applied longitudinal field ${\cal}{E} $, 
parallel to the edges ($\hat x$ direction), here we add a 
 leakage of the edge states towards a resistive propagation away from the edges,  to justify a nonzero  $\rho _{xx} \neq 0 $ coexisting with  the 
 plateaux  of $\sigma_{xy}$ at $\nu =0$. An expression for the Hall 
voltage $V_H$,  neglecting thermal effects is derived. Eventually, the model  is indeed shown to 
reproduce some  of the conductance features of both HgTe QW and graphene Hall bar.  
However, NLR across the Hall bar is probed in a different setup, by injecting current across the Hall bar ($\hat y $ direction) at the origin of 
our $\hat x -$axis   and measuring the voltage at a point $x= D  >0 $. 
 
In Section \ref{sec4} the model of the previous Section is extended, to  include  the thermal current and a  transverse applied electric field, 
together with the Peltier thermal gradient which induces diffusion of the imbalance carriers in the bulk  
along the $\hat x$ direction, with particles spatially separated  from holes, away from the applied bias.
The diffusive  longitudinal carrier transport with velocity  $v_F$ is connected to the charge imbalance via  an additional
constitutive equation, Eq.(\ref{conQ}). Phenomenologically,  it is possible to 
include the  charge imbalance relaxation rate  in this  equation.
The diffusion equations involving  the temperature and the charge imbalance chemical potential on one side and the imbalance charge density
and the non local voltage (NLV)  
on the other,  are derived and  approximately solved for the 'near region', close to the origin. Nernst effect fixes the boundary 
condition at the origin.  Details of the derivation are given in Appendix A,B,D,E and F. The  
consistency  of the equations is discussed, as well as their physical content. 
The Fermi liquid transport parameters are discussed in Appendix C. 
  
  In Section \ref{sec5} an estimate of the magnitude of the Ettingshausen parameter, $P_E$, 
  is derived for a magnetic field  $\sim 2\: Tesla$.  The thermal conduction and  the Peltier and Nernst coefficients  are 
  chosen in the range of experimental values quoted in the graphene  literature. 
  Thermal transport necessarily involves also some thickness of the graphene Hall bar which cannot be a priori determined.
  The delicate point here is the estimate of the  three dimensional  (3d-) thermal energy per  carrier $eQ$  and the related  
  3d- effective carrier density $n_{3d}$. The bulk conductivities for particles and holes  are extracted from the model 
  of Ref.[\onlinecite{Nachawaty2018}]. The parameter choice is discussed in Appendix C.  
  The result  for $P_E$ is about one or even two order of magnitude larger than the one for bismuth. 

%  The measured Hall bars, fabricated with graphene epitaxially grown onto SiC  by chemical vapour deposition, have average longitudinal resistance $R \sim 10 \: k\Omega$. The average length of the samples  is $\ell = 400 \:\mu m $ and the average width is $ w = 100 \: \mu m$.

  In Section \ref{sec6} the non local voltage is presented, which gives rise to the NLR. The  NLV 
  is plotted in Fig.(\ref {vconb}), as a function of the magnetic field at various distances $D$ from 
  the origin and the physical picture that emerges from the plots is presented and discussed. 
 
  Section \ref{sec7} includes the Summary of the crucial points in the derivation of the results and the Conclusions.

  %which implies thermal diffusion,  is parametrized by means of a length scale for thermal diffusion $\lambda_Q$,  as $v_F \lambda_Q/\ell w$. T could include relaxation effects. 
    % if $\lambda_Q \sim 100 \:\mu m $, $\bar{T}  \sim  10^{-15} \: ^oK$ and the corresponding  imbalance potential $\mu _I $  is of the order of $nV$.
  
%  Chapter V contains few  remarks  on the basis of these  quantitative estimates. An Appendix  collects part of the derivation that is mentioned in the course of the other chapters. 

%    In the absence of phonon scattering, the transported energy  is the source of an electro- chemical potential gradient across the Hall bar in the transverse  $\hat y$ direction, which generates an ohmic resistance with the mentioned odd characteristics. 

\section{Linear response in hydrodynamics}
\label{sec2}
\setcounter{equation}{0}

In this Section the  hydrodynamical picture for matter and energy transport is reviewed, by also  taking  into account  the 
additional features of  the Dirac fluid. In particular, 
the total average charge density, expressed in terms of the  average number densities of the particle and hole 
fluids is given by  $\rho = -e ( n_e- n_h )$  ( $-e< 0$ 
is the electron charge in the following). The charge  current  density is $ {\bf J }= \rho\: {\bf u } -e\left ( {\bf \nu }_e -\nubf h\right )$, where 
 $ {\bf u} $ is the center of mass velocity of the total  fluid  and  $\nubf {e,h}$ are the fluctuations produced  by the applied electric field  $ E$ and 
 thermal gradient $\tbf$.  The particle current  density is $ {\bf J_n }= n\: {\bf u } +{\bf \nu }_e +\nubf h$,  while the energy current  density is  
 given by 
 \bea
 \jbf  _Q  &=& T\: s \: {\bf u }_Q +\mu_e \: \nubf {e} + \mu _h \nubf {h}  = (\entalp ) \: {\bf u }_Q+ \mu_I \jbf _n - \frac{ \mu}{e} \jbf . \nonumber\\
  \label{preaQ}
 \enea
In Eq.(\ref{preaQ})  $ {\bf u }_Q$ is the center of mass velocity of the  $\jbf  _Q  $ current, $s$ is the entropy per unit volume and 
  $\entalp = T\: s  + \mu_e n_e+ \mu _h n_h $ is the  enthalpy per unit volume.   ${\cal{P}}= P - M\:B$ includes the thermodynamic 
  pressure of the carrier fluid and the work done by $B$ onto the  currents induced by the  magnetization ${\bf M} (\parallel{\bf B} )$ 
  at the boundary\cite{Cooper1997}.
 
Within linear response theory, one obtains the relations: 
\bea
\left ( \begin{array}{c} \jbf \\ \jbf _ Q \end{array} \right ) = \left ( \begin{array}{cc } \hat \sigma  & \hat \alpha 
\\T \hat {\tilde\alpha} & \hat{\overline\kappa} \end{array} \right )\: \left ( \begin{array}{c} \vec{ E} \\-\vec \nabla T \end{array} \right ),
\label{mattrans}
\enea
with $\hat{\sigma}$,$\hat{\alpha }$,$ \hat{\tilde\alpha}$ and $\hat{\kappa} $ being  $2 \times 2$
matrices which depend on the coordinate, as well 
as on the  flavour  ($ e^- $ or $ h^+$) label.  

To set up  the notation, we start by considering  a one-component fluid. By definition, we get for 
the conductivity matrix 
 $\hat{\sigma} =\sigma _{xx} \hat 1 + \sigma _{xy} \hat \epsilon $, 
  where  $\hat\epsilon$  is the two-dimensional antisymmetric tensor:  $ \epsilon_{xx} = \epsilon_{yy} =0,\epsilon_{xy} =
  -\epsilon_{yx} =1$. $\hat{\alpha }$,$ \hat{\tilde\alpha}$ and $\hat{\kappa} $  are the  thermoelectric conductivities which determine
  the Peltier, Seebeck, and Nernst effects.  We will assume that, on the scale of the measured samples, the response kernels can be 
  taken in the  uniform and static limit  (by also assuming that, in performing the 
  calculations, the   $q\to 0 $ limit is taken  first, followed by  the $\omega \to 0 $ limit).   
Due to Onsager reciprocity,  
there is no difference between $ \hat{\tilde\alpha}$ and $ \hat{\alpha}$ as long as currents are uniform ( $q=0 $ limit ).
  
   The thermal conductivity, $\hat{\kappa} $, is defined as the heat current response to $-\nabla  T $,  in the absence of 
   electric current     ($ \jbf = 0 $, i.e.  electrically isolated boundaries)  and is given by 
\bea
\hat{\kappa}  = \hat{\overline\kappa} - T \hat{\tilde\alpha}  \sigma ^{-1}  \hat{\alpha} 
\label{kai}
\enea
 Instead, $\hat{\overline\kappa} $ applies to samples connected to conducting leads, allowing for a stationary current flow.
 % If the center of mass of particles and holes coincides, $ {\bf u }_Q = {\bf u }$ 

     In our case,  the relativistic stress energy tensor  provides  the energy flux $ T^0_\alpha $ from the four-momentum conservation. 
    We write the energy current for a Fermi liquid as follows, to match with the expected form from Eq.(\ref{mattrans})\cite{Hartnoll2007}: 
\begin{widetext}
  \bea
 \jbf_Q \: = \frac{\entalp}{n} \left [ \jbf - \frac{\hat \sigma }{e^2}\left \{  -T{ \vec{\nabla} }\left (   \frac{\mu}{T}\right ) + 
 \frac{e}{c}\vec{v}\times \vec B \right \} \right ]  \hspace{7cm}\nonumber\\
 \to  \frac{\entalp}{n}  \jbf  - \left ( \frac{\entalp }{n e  } \right )^2 \hat{\sigma } \: \frac{1}{T} (- \vec{\nabla}  T )+  
 \frac{\entalp}{n \: e }\:\frac{ \sigma_0 }{\sigma _B} \: \hat{\sigma} \: \vec{E}
-  \frac{(\entalp )}{n^2}\frac{\hat{\sigma }}{e^2}\: \vec{\nabla}{\cal{P}}, 
\label{jq}\\
\hat{\sigma }= \frac{\sigma_0}{1+ (\omega_c \tau)^2}\left ( \begin{array}{cc} 1 & -\omega_c \tau \\
 \omega_c \tau & 1 \end{array} \right ), \:\:\:   \hat{\sigma}^{-1}= \sigma_0^{-1} \left (\begin{array}{cc} 1 &\omega_c \tau \\
 -\omega_c \tau & 1 \end{array} \right ),  \:\:\:  \sigma_0  \sim \sigma_{3d}= \frac{\rho\: v_F \tau}{\hbar k_F}.
\label{jqsigma}
   \enea
   \end{widetext}
    $\sigma _{3d} $ is an appropriate scalar   reference conductivity of the $3d$ system assumed to be uniform and  
    $\omega_c=eB/(\hbar c k_F/v_F)$ is the cyclotron frequency for the linear energy dispersion. (We have interpreted 
    $\rho v_y$ as corresponding to $  \sigma _{yx } E_x $  for the $x$ component of the energy current $ e\: \vec{v}\times \vec B $, where 
    $ \sigma _{yx }  = \sigma _B \equiv   {\rho\: c}/B$, with reference to a Drude metal in the Hall configuration.) 
        
  The first term describes a convective flux  which corresponds  to the last term in Eq.(\ref{preaQ}). In the case of a two-component 
  fluid the first term  at the right hand side of Eq.(\ref{preaQ}), referring to the center of mass convective motion of 
  the two components, should also appear.  The energy flux  also includes a  term depending on the  derivatives of ${\cal{P}} $ 
  that is absent  in  the non relativistic result \cite{Landau}.
     However,  we recognize here  that the $x$ component  $ \partial _x{\cal{P}} \propto \vec{\nabla} \times \vec{M} \times \vec{B} $ 
     is a term arising from the energy contribution of the magnetization currents $ \vec{\nabla} \times \vec{M}$ flowing at 
     the boundaries of the sample\cite{Cooper1997}. In the following,  we will take care of the edge-bulk interaction 
in the Hall bar in a different way, which eventually enables us to get rid of  this term. 
     
  Consistency of Eq.s(\ref{mattrans},\ref{jq}) requires that 
 $  \left (\frac{\entalp }{n\: e}\right )_{3d}^2\frac{\sigma}{T}\:  = \overline{\kappa} _{xx}  $  and $-\left( \frac{\entalp}{n\:e}\right )_{3d}\: \sigma _B\:  
 \hat{\epsilon}  = T \hat{\tilde \alpha }$.
  In the case of a one-component plasma  for a Drude metal, $n \equiv n_{3d}$ and $\rho$ are the carrier and charge density, respectively.  
  The transverse $dc$ response
 yields the standard Hall effect, with  $\sigma _{xy}= \rho c/B$, and the transverse Peltier effect,  $ \alpha_{xy}= (\entalp )_{3d} c  /(T\: B) = s\: c /B $, 
 which can be interpreted as charge $\rho $ and entropy $ s$ density \cite{Cooper1997}, 
 drifting with the velocity $\vec{v}_D = c\: \vec{E}\times \vec{ B}/B^2 $. Hence,   the  Nernst coefficient is:
   \begin{widetext}
   \bea
       e_N = \frac{ E_y}{-\partial _xT}  =  -\left ( \sigma ^{-1} \alpha \right )_{xy} 
			= \left [\sigma ^{-1} \left ( \begin{array}{cc}0 & s\:c /B \\ -s \: c/B & 0 \end{array} \right )\right ]_{xy} 
			= \frac{1}{\sigma _0} \frac{ (\entalp )_{3d} c}{T\:  B} =\left( \frac{ \entalp }{\: n}\right )_{3d}\: \frac{1}{T} \frac{1}{\omega_c \tau},
     \label{set}
        \enea  
            \end{widetext}
    which diverges for $n_{3d} \to 0,$ but goes as $1/B $ at large $B$ ($\omega _c \tau >>1$)\cite{Mueller2008b}.
   When $ T >> \mu$, by posing $ \left ((\entalp )/n\right )_{3d} = k_B T $ in  the definition of  $ \overline{\kappa} _{xx}$, we get a rewriting  of  
   the Wiedemann-Franz law, to be compared with the Fermi liquid result:
   \bea
   \frac{\overline{\kappa}_{xx}}{\sigma T } = \frac{\pi ^2}{3} \left (\frac{k_B}{e} \right )^2 .
   \label{WFranz}
   \enea 
%Naively, one would conclude that the left hand side diverges when  $n_{3d} \to 0 $, what leads to a contradiction.  However, this  Fermi liquid limit does not apply to our case. 

 In a clean relativistic system at $B\neq 0 $,  Lorenz invariance implies 
 $  \sigma _{xx} (\omega  \to 0) =  \alpha_{xx} (\omega \to 0)  =0$ for a one-component plasma\cite{Mueller2008a}.
 In a reference frame moving at the constant velocity  $\vec{v}_D $ with respect to the laboratory frame, 
 the observed electric field vanishes and, hence, in that frame the charge currents vanish. As $ v_D \perp E$, 
 transforming back to the  laboratory frame, the longitudinal field is still vanishing.  
 These results hold beyond the hydrodynamic description  even when  $\omega _c \tau_{ee} >> 1$, as long as Lorentz invariance holds.

%The equality of Eq.(\ref{kai}), $ \hat{\kappa} ( \omega \to 0,\mu \to 0 ) =  - T \hat{\alpha}\hat{\sigma }^{-1} \hat{\alpha } $, follows.   

 % It follows that the  Nernst coefficient is  $ e_N = - (\hat \sigma ^{-1} \hat \alpha )_{xy}$.

% \begin{widetext}
  %    \end{widetext}

 % The first term is absent because the circuit is open. A term analogous to the second one  involving a temperature gradient component in the  $\hat{y}$ direction has been dropped. 

In graphene, electric conductivity  reaches the minimum  value \cite{Ludwig1994,Katsnelson2006} $\sigma _Q = 4 e^2/\pi  h$ at the CNP  
and  $ \overline{\kappa} _{xx} $ is finite.  Using Eq.(\ref{kai}),  we get \cite{Mueller2008a}:
  %  \begin{widetext}
    \bea 
   \kappa _{xx}^{FL} ( \omega =0,\mu )  =
  % =- \left ( \begin{array}{cc } 0  &\alpha _{xy}  \\ -\alpha _{xy}  & 0 \end{array} \right )\:  \left ( \begin{array}{cc } 0  &-\sigma _{xy}  \\ \sigma _{xy}  & 0\end{array} \right )\: \frac{1}{\sigma _{xy}^2 + \sigma _{ xx}^2}    \left ( \begin{array}{cc } 0  &\alpha _{xy}  \\ -\alpha _{xy}  & 0 \end{array} \right )\:
 \left (  \frac{\entalp }{n	\: e}  \right )_{3d}^2\: \frac{\sigma _{3d}}{T}\left [ 1 - \left (\frac{\sigma _B }{\sigma _{3d} }\right )^2 \right ]. \nonumber\\
     \label{seit} 
   \enea
Similarly, 
\beq
\alpha_{xy}^{FL} =  \frac{  \left(\epsilon + {\cal{P}}\right )_ {3d} }{T\: B}c.
\label{seit2}
\eneq
%\end{widetext}
\begin{figure}
\begin{center}
\includegraphics[width= 0.7 \linewidth]{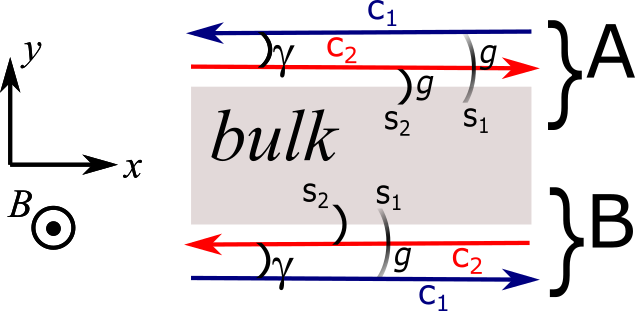}
\caption{(a)  Transport in the Hall bar geometry.  The edge states are denoted by red and blue lines and 
carry opposite charge carriers. The various couplings between the edge states and the bulk are also indicated. 
The notation used in the text for  the edge chemical potentials $c_i $ ($ i=1,2$) and bulk  chemical potentials 
$ s_i$  is reported.}
\label{scheme}
\end{center}
\end{figure}

Moving now to the two component  fluid for a uniform isotropic system, with  average  $3d$ carrier number density  $n_{3d} = n_e+n_h $,
a derivation similar to the one given in Eq.(\ref{jq}) provides   contributions to the  fluctuations 
$\nu _{e,h}$ appearing in  
the charge and particle current densities $ \jbf  $ and $ \jbf  _n$, in terms of  the gradient of the  temperature,
of  $ \mu$,  and of $ \mu_I$,  by means of the conductivities $ \sigma _{ee}, \sigma _{hh}$ and $\sigma _{eh}$.
At zero applied magnetic field, the conductivities are constructed from the Drude relaxation times $ \tau_{ee}, \tau_{hh}$ and $\tau_{eh}$ 
as\cite{Foster2009}:
\begin{widetext}
 \bea
e\: \left ({\bf J}- \rho {\bf u}\right) = - e^2\:\left (   {\bf \nu} _e  - {\bf \nu} _h \right ) = [ \sigma _{ee} +\sigma _{hh}-2\sigma _{eh} ] 
\left ( e {\bf E} - T  {\bf \nabla}  \frac{\mu}{T}  \right )  +  [ \sigma _{ee} -\sigma _{hh}  ]   \left ( - T {\bf \nabla}  \frac{\mu_I}{T} \right )\nonumber\\  
-e^2\: \left ({\bf J}_n- n {\bf u}\right) = - e^2\:\left (    {\bf \nu} _e  + {\bf \nu} _h \right ) = ( \sigma _{ee} -\sigma _{hh} ) \left 
( e {\bf E }- T {\bf \nabla}  \frac{\mu}{T}  \right )+  [ \sigma _{ee} +\sigma _{hh}+2\sigma _{eh} ] \: \left (  - T {\bf \nabla}  \frac{\mu_I}{T}  \right ). 
   \label{ali}
      \enea
      \end{widetext}
While $\hat{\sigma }_{ee}$ and $\hat{\sigma} _{hh}$ in Eqs.(\ref{ali}) are easily interpreted as particle and hole conductivities, 
respectively (in the next Sections they will be generically referred to as $\sigma _{ab, i}$, with  $ a,b =   x,y$ and    $i=1,2$, 
respectively for  $e$ and $h$), $\hat{\sigma} _{eh} =\hat{\sigma} _{he}$ refers to a drag conductivity between particles
and holes. $\sigma _{ee} +\sigma _{hh}-2\sigma _{eh}$ plays the role of $\sigma _{3d}$\cite{Foster2009}. 
We consider the limit in which both $\hat x -$component terms of  $ \jbf  $ are negligibly small 
(that is  $ E_x \approx \partial _x \mu \approx 0 $ and to linear order in $\hat{\sigma} _{ee} - \hat{\sigma} _{hh}$, as $\mu _I 
\propto \hat{\sigma} _{ee} - \hat{\sigma} _{hh}$, as well) and the only contribution relevant to our derivation  is the  $x-$component of 
the second term at the r.h.s. of  the particle current  $ \jbf  _n$, of dimension $[ energy / t \ell ^2] $. As 
this is the most important contribution appearing in our model, in the following we 
extensively discuss about it.
 
While Hall transport is essentially  a $2d$  phenomenon, particle current and  thermal transport are essentially   $3d$. Accordingly 
in the following we and we will have to carefully account for that difference.  Also, defining the  conductivities requires here some care. 
Here, $\sigma _i\equiv \sigma _{xx,i }$ and $ \sigma _{xy,i}  ( i=1,2) $  are $3d$
conductivities and have dimension $[1/t]$. Also, an aspect ratio must be  be introduced to take into account 
the effective  thickness $d \approx  0.3\: \AA  $ of the graphene sheet grown on top of the $SiC$ 
surface in the Hall bar,  so that the $ \sigma _{ab,i} $ conductivities are related to the  longitudinal 
resistance $R$ according to:
                   \bea
                   \frac{1}{\sigma _{3d} }  = R \frac{ \ell d}{w}.  
                   \label{resa}
                   \enea
                   This implies that  a  volume unity  $ {\cal{V}} \propto  [\ell w d]$ can be  introduced,  which will appear in the rest of the paper. 
                   
 As mentioned in the Introduction, the particle and energy  currents along the $\hat x$ direction are related to the imbalance between electrons and 
 holes induced  by the applied electric field ${\cal{E}} \: \hat y$ and to the corresponding electric current $ J_y$.  We are mostly interested  
 in  the diffusion  of the $\hat {x}-$component of $ \jbf  _n$ along the $\hat x$ direction. Here we introduce 
 $\delta  \jbf  _n \equiv \jbf  _{n,A}-  \jbf  _{n,B}$, where A,B denote areas close to each of the two edges, which  is not a
 locally  conserved current.   In a steady state, the  diffusion  $ \partial_x  \delta J_n   $  of the particle   current density 
 is induced by the  particle/hole  imbalance charge $\rho_I$, which we will appropriately define in Section  IV, Eq.(\ref{imbo}),
 in terms of the chemical potential imbalance. Accordingly,  we write: 
   \beq
   \frac{ \partial }{\partial x} \left ({\bf J}_{n}\cdot \hat{x} \right )  = -   v_F \frac{ 1}{n_\square  {\cal{V}} }  \: \frac{  \rho _I }{e} \: .
   \label{conQ}
   \eneq
Here $n_\square {\cal{V}}$ plays the role of a  phenomenological effective  relaxation  length for the  thermal diffusion $\lambda _Q$, 
which is  $ \sim  \mu m$. (Phonon scattering is expected to play no role as it  is virtually frozen out  at low temperature.)   

   On the basis of the previous remarks, we are eventually able to build up a possible interpretations of  the non local
   resistance of Fig.\ref{tef23}, by assume an open circuit in the $\hat x$ direction parallel to the Hall bar edges.  
   On injecting a current across the Hall bar at $x= 0$, a non-equilibrium charge imbalance, as well 
   as a temperature gradient $ -\partial _x T (x) $, are generated  in the $\hat x$ direction, due to the  Ettingshausen effect.    
   Enhanced thermal conduction $ \kappa _{xx} (x)$ along the Hall bar drives the particle current, with space separated
   particle and holes close to edges $A$ and $B$  so that the imbalance   reaches the point $x = x_b$ where a transverse 
   voltage $v_y$ is measured.  Charge conduction in the edge channels  is assumed not to be influenced by thermal processes, 
   but they provide some dissipation and a  leakage into ``bulk'' states  of carriers of opposite charge at  opposite edge states. 
  
 \section{The Abanin {\it et al.} model} 
\label{sec3}
  \setcounter{equation}{0}
    In this Section we recall   the  main features of the model of Ref.[\onlinecite{Abanin2007}]. 
While the model includes the longitudinal dissipation coexisting with the leakage of carriers from the edges into 
    the bulk of the Hall bar, it  does not include thermal effects. However, its careful discussion  
is an important preliminary step for us, in order to extend it by including thermal effects, as well, 
which will be the subject of  Section IV. 
    
 Under the action of  an orthogonal  magnetic field $B$, close to the  CNP ($\mu \approx 0 $),  particle and holes states 
 counterpropagate  at the edges of the  graphene Hall bar. Let  $\hat x$ be the direction along the edges and  $\hat y$
 the direction orthogonal to the edges and $w$ be the width of the Hall bar. 
 We denote by  A the upper edge at $y=0$   and  by  B the lower edge   at $y=w$.
 In the following, labels A,B refer to the  region close to the edge A and B, respectively,
 while the interior of the Hall bar will be denoted as the "bulk" (see Fig.(\ref{scheme})). 
 Charge transport  along the edges   is described by the electrochemical potentials $c_{1,2} $ for 
 particles and holes, respectively. (Here, we do not account for the spin, which merely provides a factor of 2 in the final 
 results). As the Hall bar is not fully insulating in the bulk, we assume nonzero isotropic bulk conductivities  
 $\sigma _{xxi}=\sigma _{yyi} = \sigma _i$ (the label $i=1,2$ is for $e$ and $h$  respectively) so that,  
 if $\psi_{1,2} (y) $  are the electrochemical potentials for the bulk,  $\sigma _{yy i} \:\hat y\cdot \nabla \psi _i (y)$ 
 is  the bulk current  orthogonal to the edges.    The terms describing leakage of current from the  edges into the bulk are  
respectively given by  $g' (\psi_i - c _i)_A$ and $ g' (c_i- \psi_i )_B$. At edge A the  linear response to an   
electric field ${\cal{E}}$  in the $\hat x$ direction is ${\cal{E}}  = \gamma' ( c_1-c_2)_A + ( - )^i\: g' ( \psi_i ( y=0) -c_i)_A $, 
while, at edge B, ${\cal{E}}  = -\gamma' ( c_1-c_2)_B - ( - )^i\: g' ( \psi_i ( y=w) -c_i)_B $.  The electrochemical potentials 
$c_{1,2} $ and $\psi_{1,2} (y) $ are complementary to the density fluctuations $\nu _{e,h}$ of the previous Section. 
  In the absence of thermal effects,  we assume mirror symmetry   w.r.to the longitudinal axis lying halfway between edges A and B, that is, 
we set $\psi_i(0) = -\psi_i(w) = s_i$. The derivative  of the bulk electrochemical potential in the  $\hat {y}-$direction is 
linearized as  a  finite difference $\hat {y}\cdot  \nabla  \psi _i \sim (\psi_i(0)  -\psi_i(w))/w= 2\: s_i /w $. 
 The electric field $\ebf$ is in the $\hat{x}-$direction and appears in the four equations for the edges  ($ 1e,2e,3e,4e$):
  \bea
   1e:\:\:\: & -{\cal{E}} & = -\gamma' ( c_1-c_2)_A + g' ( s_1-c_1)_A \nonumber \\
    2e:\:\:\: & {\cal{E}} & = \gamma' ( c_1-c_2)_A + g' ( s_2-c_2)_A \nonumber \\
       3e:\:\:\: & {\cal{E}} & = -\gamma' ( c_1-c_2)_B - g' ( s_1+c_1)_B \nonumber \\
        4e:\:\:\: & -{\cal{E}} & = \gamma' ( c_1-c_2)_B - g' ( s_2+ c_2)_B 
        \label{ureqe}
        \enea
The bulk currents explicitly appear in the four equations for the bulk  ($ 1b,2b,3b,4b$): 
        \bea
         1b:\:\:\: & 0 & = -2\frac{\sigma_1}{w}\:  s_1 +\sigma _{xy,1}  {\cal{E}} + g ( c_1-s_1)_A \nonumber \\
            2b:\:\:\: &  0 & = 2\frac{\sigma_2}{w}\: s_2 -\sigma _{xy,2}  {\cal{E}} - g ( c_2-s_2)_A \nonumber \\
              3b:\:\:\:  & 0 & = 2\frac{\sigma_1}{w}\: s_1 -\sigma _{xy,1}   {\cal{E}} + g ( s_1+c_1)_B \nonumber \\
                 4b:\:\:\: & 0 & = -2\frac{\sigma_2}{w}\: s_2 +\sigma _{xy,2}   {\cal{E}} - g ( s_2+c_2)_B 
                 \label{ureq2}
                 \enea 
                    %             $d \sim  1 \AA$  is introduced to define an 'unity' for the  $3d$ average density, according to $ \delta n_{3d} = [ \ell w d]^{-1} $. 
The conductivities $\gamma ', \gamma $ and $g',g$ differ just by unities, as the former (primed) ones  have
dimension $[1/L]$  while the latter ones  have dimension $[(L\: t)^{-1}]$. Accordingly,  the chemical potentials  
have dimension $[e/L]$ and the the electric field $ {\cal{E}}$ has dimension $[e/L^2]$. 

Note that,  the sum of Eq.s(\ref{ureqe}:1e,2e) minus  the sum of Eq.s(\ref{ureqe}:3e,4e) yields:
                 \bea
               \frac{1}{2} \left [   ( c_1 +c_2 )_A - ( c_1 +c_2 )_B  \right ]  =   ( s_1+ s_2),  
                  \label{vh}
                  \enea
                  where, at the right hand side, we write $ \left [ ( s_1+s_2 )_A + (s_1+s_2)_B  \right ]/2 =  ( s_1+ s_2)  \equiv  s_+ $.
                  This is consistent with isotropy in the bulk.  This approximation holds  approximately also when thermal effects are taken
                  into account,  as long as  the relaxation length for the carrier imbalance $\lambda _Q$ is large enough with respect to the width 
                  $w$ and the length $\ell $ of the Hall bar. 
                 The left hand side  is  the definition of the Hall voltage $ V_H $.   In the absence of  thermal effects the Hall voltage  
                 can also be defined from the bulk potentials as   $ V_H  = s_+$. The  corresponding electrochemical potentials $\mu$, 
                 being charge dependent,  involve the differences instead of the sums:
                 $c_{-,A}= (c_1-c_2)_A$,  $c_{-,B} = (c_1-c_2)_B$, $s_- = (s_1-s_2)$.   
                 The equality  in Eq.(\ref{vh}) states the absence of the carrier  imbalance between edges and bulk, which 
 is a typical situation in   the absence of thermal effects. Indeed, in the presence of thermal  effects (which we take into 
 account in the next Section by assuming that they play a role in the $3d-$like bulk  but not  at the edges),
 it is indeed violated and, as it  will clearly appear in the following, 
 the  non-equilibrium difference $ ( c_1 +c_2 )_A - ( c_1 +c_2 )_B -2( s_1+ s_2) \equiv  c_{+A}-c_{+B}- 2\: s_+$ provides the imbalance  chemical 
 potential $\mu _I$  close to the  edges A and B.  
                  
\begin{figure}
\begin{center}
\includegraphics[width= 0.7 \linewidth]{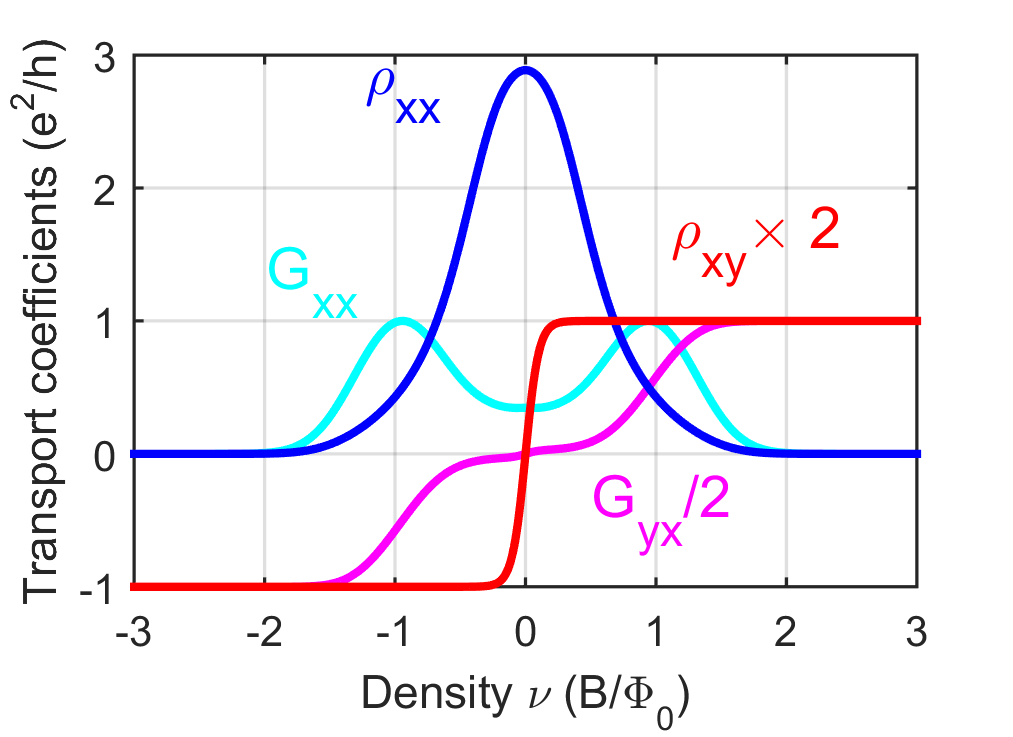}
\caption{Results of the Abanin model,  reproduced from Ref.[\onlinecite{Abanin2007}]. 
Density dependence of the transport coefficient $\rho_{xx}$, $\rho_{xy}$, $G_{xx}$= $\rho_{xx}/(\rho_{xx}^2+\rho_{xy}^2)$, 
$G_{yx}$ = $\rho_{xy}/(\rho_{xx}^2+\rho_{xy}^2)$. The conductivities are modeled by Gaussians centered at 
$\nu \pm 1$, $\sigma_{xx}^{(1,2)}$= $\exp(-A(\nu\pm 1)^2)$ and $\sigma_{xy}^{(1,2)}$ are given by the 
semicircle relation.  $A=5$, $gw=1$ and $\gamma w$ = 6 are used.}
\label{scheme2}
\end{center}
\end{figure}

                       \begin{widetext}
              We now  choose  as independent variables: $c_{-,A}$,  $c_{-,B}$, $s_1,s_2$ and drop the apex in the definitions of 
              the conductivities, with an appropriate choice of the units for $\ebf$.   From Eq.s (\ref{ureq2})
              we get ( $ \sigma _{xy,i} \equiv \eta _i $): 
\bea
 \left ( \begin{array}{cccc} -(2 \gamma +g)  & 0 & g & -g \\ 0 & -(2 \gamma +g)  & -g & g \\ \gamma & 0 & 0 &  -2 \frac{\sigma_2}{w} \\ 0
 & -\gamma  &  2 \frac{\sigma_1}{w} & 0 \end{array} \right )\:  \left ( \begin{array}{c}  c_{-,A} \\ c_{-,B} \\  s_1 \\ s_2 \end{array} \right )= 
 \left ( \begin{array}{c} -2\:{\cal{E}}  \\ 2\:{\cal{E}} \\  - ( \eta _2 -1) \: {\cal{E}}  \\  ( \eta _1 +1) \: {\cal{E}}  \end{array} \right )
 \enea
   The solution is:
   \bea
   (c_1-c_2)_A =  - (c_1-c_2)_B = \left [ 4 + wg\left ( \frac{1}{\sigma_1} +\frac{1}{\sigma_2}  \right ) - wg \left (  \frac{\eta_2}{\sigma_2} -
   \frac{\eta _1}{\sigma_1}  \right ) \right ] \cdot  {\cal{E}}/ D \nonumber\\
   D= 2 ( 2\gamma + g) + gw\gamma \left ( \frac{1}{\sigma_1} +\frac{1}{\sigma_2}  \right )  \nonumber\\
   s_1 =  \left [ \gamma g ( \eta_1 + \eta_2 ) + 2 ( 2\gamma +g) \: \eta _1 \sigma _2  + 2g \sigma _2 \right ] \cdot  {\cal{E}}/ ( 2 \sigma _1\sigma _2 D )  
   \nonumber\\
    s_2 =  \left [ \gamma g ( \eta_1 + \eta_2 ) + 2 ( 2\gamma +g) \: \eta _2 \sigma _1  - 2g \sigma _1 \right ] \cdot  {\cal{E}}/ ( 2 \sigma _1\sigma _2 D )  
   \nonumber\\
   \label{eqsol}
\enea
    
   The Hall Voltage of Eq.(\ref{vh})  is readily  obtained as:
   \bea
    V_H =  -  2 w ( 2\gamma +g) \: \left [ \tilde\eta _1 \left ( \sigma _2  +\lambda  \right ) +   \tilde\eta _2
    \left ( \sigma _1  +\lambda  \right )  \right ] \cdot  {\cal{E}}/ ( 2 \sigma _1\sigma _2 D ), 
    \label{rifa}
    \enea
    \end{widetext}
    where  $\lambda =  {g \gamma w }/{(2\gamma +g)} , \tilde\eta _1 = \eta _1 + g/( 2\gamma +g) $ and  $ \tilde\eta _2 = \eta _2- g/( 2\gamma +g) $.
    The  ratio $g/(2\gamma +g) $, plays the role of a contribution to the Hall conductivity  determined by current  leaking from the edges into the bulk.
Eq.s(\ref{rifa}) were originally  derived in  Ref. [\onlinecite{Abanin2007}]. For completeness, here we report  the plots of  the longitudinal 
      and transverse resistivity $\rho _{xx},\rho _{xy}$, together with the corresponding conductances $G _{xx}$ and $G _{xy}$ as 
      functions of the filling $ \nu $ (Fig.(\ref{scheme2}))\cite{Abanin2007} . A kind of plateau at $ \nu \approx 0 $ is recognizable in 
      $G _{xy}$  accompanied by a large peak in $\rho _{xx}$.

  \section{Thermal relaxation of  charge imbalance} 
  \label{sec4}
  \setcounter{equation}{0} 
  
Here we extend the model of Section III to include  energy transport  and  thermal  effects on the conductance. To keep in touch 
with the experimental setup we discuss, we consider  the Hall  bar  of length $\ell$   as an open circuit in the longitudinal ($\hat{x}$) 
direction.    A current is  injected  in the $\hat y $ 
direction, orthogonal to the edges A and B,  by  applying  an  electric field ${\cal{E}}$ at the  contacts at $x= 0 $.  We assume 
that no thermal effects involve the edges.  Therefore  Eq.s(\ref{ureqe},1e-4e) for the edge propagation in the $\hat{x}-$direction 
do not change, except for the fact that  there is  no driving electric field at open circuit. However,  the electrochemical potentials
are expected to depend on the $\hat{x}$ coordinate along the edges and their derivative replaces the external electric field that was 
applied in the model of Section III, when the circuit was closed. Therefore, the equations for the edges are now 
given by : 

 \bea
   1e & :\:\:\:  \partial _x c_{1A}  & = -\gamma' ( c_1-c_2)_A + g' ( s_1-c_1)_A \nonumber \\
    2e & :\:\:\: - \partial _x c_{2A} & = \gamma' ( c_1-c_2)_A + g' ( s_2-c_2)_A \nonumber \\
       3e &:\:\:\: - \partial _x c_{1B}  & = -\gamma' ( c_1-c_2)_B - g' ( s_1+c_1) _B\nonumber \\
        4e & :\:\:\:  \partial _x c_{2B} & = \gamma '( c_1-c_2)_B - g' ( s_2+c_2)_B 
        \label{typea}
        \enea
  We now look for the components of ${\bf J}$  and of  ${\bf J}_Q$ in  Eq.s (\ref{preaQ}) respectively 
   along  the  $ \hat y $ direction and   along  the  $ \hat x $ direction. Eventually, on inserting them   
 in Eq.(\ref{mattrans}), we trade it for a system of  differential equations for the  
  temperatures and the chemical potentials  only.                                    
                       
    Close to the edge A, or B, the components of   ${\bf J}$  in  the  $ \hat y $ direction, coming from opposite edges,  
    can be recovered by respectively summing  Eq.s(\ref{ureq2}: 1b,2b )  and  Eq.s(\ref{ureq2}: 3b,4b ). 
    The terms $\sigma _{yx,i} {\cal{E}}_x $    appearing in  Eq.s(\ref{ureq2}: 1b-4b)  vanish.  
    The portion of charge current lost by the edge potential $(c_1 -c_2)_A$ is, 
    to first order in $g$, the fraction of current  emerging from the imbalance close to the edge, $ \frac{g}{2\gamma+g} t_0^{-1}  
    \partial _x ( c_1+c_2)_A $, as derived from the difference between  Eq.s(\ref{typea}, 1e,2e).  $t_0$ is an appropriate time scale which 
    we will not have to specify in the following.  The terms $ \propto g $  should not be counted twice.        
    The corresponding manipulations apply in the region close to edge B. We therefore set: 
           \begin{widetext} 
    \bea
        1b':\:\:\:   J_A \cdot   \hat y - \rho \: u  = - ( \nu _e  -\nu _h )_A  & =  -2 \left ( \frac{\sigma_1}{w}\: 
        s_1 -\frac{\sigma_2}{w}\: s_2  \right ) _A + g ( c_1 -c_2 )_A - g\: ( s_1-s_2 )_A 
        +\frac{g}{2\gamma +g} \: t_0^{-1}\partial _x ( c_1+c_2)_A\nonumber \\
            3b':\:\:\:   J_B \cdot   \hat y + \rho \: u =  ( \nu _e  -\nu _h )_B & =    2 \left ( \frac{\sigma_1}{w}\:  
            s_1 -\frac{\sigma_2}{w}\: s_2  \right )_B - g ( c_1 -c_2 )_B - g \: ( s_1-s_2 )_B
            +\frac{g}{2\gamma +g} \:t_0^{-1} \partial _x ( c_1+c_2)_B. \nonumber \\
                           \label{1e3bp}
                 \enea
                  \end{widetext} 
                  
Due to the  open circuit condition along $\hat{x}$, ${\bf u}$ is in  the $\hat y $ direction, as a consequence of  the applied 
electric field ${\cal{E}} \hat y$.
          
 On the contrary, the components of the current  ${\bf J}_n$, which give  the fluctuations in the carrier  transport  along the $\hat x$ direction,
 can be induced from the difference between Eq.s(\ref{ureq2}: 1b,2b) for area A, or  Eq.s(\ref{ureq2},  3b,4b)for area B. 
 The extra  term  to be considered coming from  Eq.s(\ref{typea}:1e,2e) for edge A is given by   $ t_0^{-1}  \partial _x ( c_1-c_2)_A $. 
 The same happens for region B.  Therefore, we get: 
            \begin{widetext} 
 \bea
                    2b':\:\:\: \left ( {\bf J}_{n} - n\: {\bf u } \right )_A \cdot   \hat x =- ( \nu _e  +\nu _h )_A & = -2 \left ( \frac{\sigma_1}{w}\: 
                    s_1 +\frac{\sigma_2}{w}\: s_2  \right )_A  +  g ( c_1 +c_2 )_A -g ( s_1+s_2 )_A + t_0^{-1}\partial _x ( c_1-c_2)_A \nonumber \\
            4b':\:\:\:  \left ( {\bf J}_{n} - n\: {\bf u } \right )_B \cdot   \hat x = ( \nu _e  +\nu _h )_B & = 2 \left ( \frac{\sigma_1}{w}\:  
            s_1 +\frac{\sigma_2}{w}\: s_2  \right )_B - g ( c_1 +c_2 )_B -g ( s_1+s_2 )_B+ t_0^{-1} \partial _x ( c_1-c_2) _B
                          \label{2e4bp}
                 \enea
                 \end{widetext} 
             (we have reported the labels $1b'-4b'$ on the very left, in correspondence with the ones of  Eq.s(\ref{ureq2})).   
              
   The  gradients of the electrochemical and imbalance chemical potentials, 
 $\nabla \mu, \nabla \mu_I $  appear here explicitly, according to the identification  $  \partial_y \mu = \partial _y 
 ( \mu _e - \mu _h )/2 \sim[ ( c_1  -  s_1)_A - ( c_2 - s_2 )_A] /w$ and  $ \partial_y  \mu_I = \partial _y ( \mu _e + \mu _h )/2 
 \sim [ ( c_1  + c_2)_A - ( s_1 + s_2 )_A] /w $, so that Eq.(\ref{1e3bp},\ref{2e4bp}) are a special case of  Eq.s(\ref{ali}),      
 adapted  to our case.    
 
   Although  Eq.s(\ref{2e4bp}) refers to the fluctuating part of the particle current  ${\bf J}_{n} $ only, 
   they are sufficient to complete the characterization of the thermal effects from  Eq.(\ref{mattrans}).  
   This is because, according to Eq.(\ref{jq}), just linear terms in the gradients  should be added to   
   $   (\entalp ) \: {\bf u }_Q +  (\entalp ) \: {\bf J } $ to obtain $\jbf  _Q $ (see also  Eq.s(\ref{ali})).
   According to Eq.(\ref{preaQ}),   $\jbf  _Q -   (\entalp ) \: {\bf u }_Q -  (\entalp ) \: {\bf J }  \approx  \mu_I \jbf _n $,
   if we approximate $\mu /e$ with $(\entalp )/n$.  This implies that it is enough to know   
   the contributions  $ - e \: [ \nu _e  -\nu _h ] $  and $ -   [ \nu _e  +\nu _h ] $ to  the charge current density and  
   to the number current density, respectively,    to recover informations about the fluctuating components,  which are 
   proportional to the  gradients of the temperature and of the chemical potential. 
 \\

  \subsection{Response to thermal and potential gradients }
  % This odd choice allows to assign equal units to the tensors  $\hat{\alpha} $ and $\hat{\kappa}$ $\propto [ \ell \: T^{-1}] $  in the rest of the paper.  
    
                        Following our derivation above, we now derive the equations describing the 
bulk charge and particle currents arising in response to applied electric field  and thermal gradients, 
 ${\cal{E}}$  and $\nabla T$,  via  the electrical and thermal conductivities  corresponding  to Eq.s(\ref{mattrans}),
                        by means of Eq.s(\ref{1e3bp}, \ref{2e4bp}). 
In our geometric, the electric field is orthogonal to the edges, while the thermal gradient is parallel. In the reference frame 
in which ${\bf u}_Q = 0 $,  we get the following Eq.s(\ref{ureqt}: 1b',3b') 
                        for the charge current oriented toward  the bulk and orthogonal to the edges, and the following Eq.s(\ref{ureqt}: 2b',4b') for 
                        the  energy current in the $\hat x $ direction, produced by particle-hole imbalance:  
 \begin{widetext}
          \bea
        1b': & \:\:\:  -\tilde \sigma   {\cal{E}} +\alpha _{yx} ( -\partial _x T )_A  & = - \rho \: u -2 \left ( \frac{\sigma_1}{w}\:  s_1 -
        \frac{\sigma_2}{w}\: s_2  \right )_A  + g ( c_1 -c_2 )_A - g \: ( s_1-s_2 )_A 
        +\frac{g}{2\gamma +g} \: t_0^{-1} \partial _x ( c_1+c_2)_A\nonumber \\
            2b': & \:\:\: T_A\: \alpha _{xy} {\cal{E}} +\kappa_{xx} ( -\partial _x T )_A  & = \left ( \frac{\epsilon + {\cal{P}} }{ n\: e}
            \right  )_{3d}\left [-2 \left ( \frac{\sigma_1}{w}\:  s_1 +\frac{\sigma_2}{w}\: s_2  \right )_A  +  g ( c_1 +c_2 )_A -g ( s_1+s_2 ) 
            +t_0^{-1} \partial _x ( c_1-c_2)_B\right ] \nonumber \\
            3b': &  \:\:\:  \tilde \sigma   {\cal{E}} - \alpha _{yx} ( -\partial _x T )_B  & =  \rho \: u + 2 \left ( \frac{\sigma_1}{w}\: 
            s_1 -\frac{\sigma_2}{w}\: s_2  \right )_B - g ( c_1 -c_2 )_B - g \: ( s_1-s_2 )_B
            +\frac{g}{2\gamma +g} \: t_0^{-1} \partial _x ( c_1+c_2)_B \nonumber \\
            4b': & \:\:\:  - T_B \:\alpha _{xy} {\cal{E}} - \kappa_{xx} ( -\partial _x T )_B  & =   \left ( \frac{\epsilon + {\cal{P}} }{ n\: e}
            \right  )_{3d}  \left[2 \left ( \frac{\sigma_1}{w}\:  s_1 +\frac{\sigma_2}{w}\: s_2  \right )_B - g ( c_1 +c_2 )_B -g ( s_1+s_2 )+t_0^{-1} 
            \partial _x ( c_1-c_2)_B\right ]. \nonumber\\
                          \label{ureqt}
                 \enea
                 
                         \end{widetext}
    (Note that, from now on, we refer all the physical quantities to a 3-d system so that, e.g., the conductivities have dimension
                         $sec^{-1}$, according to Eq.(\ref{resa})).
                         
When comparing  Eq.s(\ref{ureqt}) with   Eq.s(\ref{ureq2}), we see that  now the terms at the left hand side  of  
                         Eq.s(\ref{ureqt})  are no longer equal to 0, due to the  electric field   ${\cal{E}} \neq 0$ (in 
                         the $\hat y$ direction) and to the coupling to the thermal gradient, according to Eq.(\ref{mattrans}). 
Also, note that, in Eq.s(\ref{ureqt}: 2b',4b'),                         
we have introduced the prefactor $  \left ( \frac{\epsilon + {\cal{P}} }{ n\: e}\right  )_{3d} $, 
     to account for  the proportionality in Eq.(\ref{jq}) and  in                   
           Eq.(\ref{2e4bp}).     
                          
Within linear approximation, we assume that Eq.s(\ref{typea}:1e-4e) still hold to lowest order and we use them to substitute the derivative
term $ \partial _x ( c_1+c_2)_A $  and $ \partial _x ( c_1-c_2)_A $ from  Eq.s(\ref{typea}:1e, 2e)  into Eq.s(\ref{ureqt}: 1b') and  Eq.s(\ref{ureqt}: 2b') 
respectively, and similar derivative terms  from   Eq.s(\ref{typea}:3e, 4e) into  Eq.s(\ref{ureqt}: 3b') and  Eq.s(\ref{ureqt}: 4b') respectively,  to get: 
 
  \begin{widetext}
   \bea
    1b':\:\:\:  -\tilde \sigma \: {\cal{E}} +\alpha _{yx} ( -\partial _x T )_A  & =- \rho \: u  -2 \left ( \frac{\sigma_1}{w}\:  s_1 -
    \frac{\sigma_2}{w}\: s_2  \right )_A  - g\: \frac{2\gamma}{2\gamma+g }\: s _{-,A}\nonumber \\
            2b':\:\:\: T_A \:  \alpha _{xy} {\cal{E}} +\kappa_{xx} ( -\partial _x T )_A &  = -2\:  \left (\frac{ \epsilon + {\cal{P}}  }{ n e}
            \right )_{3d} \left ( \frac{\sigma_1}{w}\:  s_1 +\frac{\sigma_2}{w}\: s_2  \right )_A  \nonumber \\
            3b':\:\:\:  \tilde \sigma \: {\cal{E}} - \alpha _{yx} ( -\partial _x T )_B  & =  \rho \: u + 2 \left ( \frac{\sigma_1}{w}\:  s_1 -
            \frac{\sigma_2}{w}\: s_2  \right )_B - g\: \frac{2\gamma}{2\gamma+g } \: s_{-,B}\nonumber \\
            4b':\:\:\:  - T_B \:  \alpha _{xy} {\cal{E}} - \kappa_{xx} ( -\partial _x T )_B  & = 2   \left (\frac{ \epsilon + {\cal{P}}  }{ n e}
            \right )_{3d} \left ( \frac{\sigma_1}{w}\:  s_1 +\frac{\sigma_2}{w}\: s_2  \right )_B .
                          \label{ureqt1}
                 \enea
                        \end{widetext}          
                         %  Here we have defined  $ s_{\pm, A/B} =  s_{1A/B}\pm s_{2A/B}$ and  analogously    $ c_{\pm, A/B} =  c_{1A/B}\pm c_{2A/B}$  will appear  in the following.  
                                                              
 Eq.s(\ref{ureqt1}) are pretty remarkable, as they express the stationary linear response  to electrical and thermal  
                          perturbations just in terms of the bull electrochemical potentials, with the coupling 
                         of the bulk   to the edges encoded  in  Eq.s(\ref{typea}).  
                                                    
 We now use   $\Delta T=T_A- T_B $ to denote the   difference 
                                                       in the fluctuations of the temperature  in the  $\hat y$  direction. The average 
                                                       temperature $\overline {T} (x)/2 = (T_A+ T_B )/2$, fluctuates  around the  
                                                       temperature of the bath $T= 1 \: ^o\! K$. 
                                                       The fluctuation within  $\overline {T} (x)$ is a small fraction of the  bath temerature.  
 Eq.s(\ref{typea},\ref{ureqt1}) determine set of 8  equations in the 9   unknowns $ \overline {T} (x) ,\: \Delta T , \: s_{-,A} \pm s_{-,B},
                                                       \: c_{-,A} \pm c_{-,B}, \: c_{+,A}, \pm c_{+,B},\: s_+$, which eventually 
yields the space dependent relaxation  along the  $\hat x$ direction, when  a  
                                                       steady state perturbation  acts at $x=0$.  
                             
                             To close the corresponding set of linear differential equations, we add to it the constitutive equation
                             in Eq.(\ref{conQ}),  which quantifies the energy flux in the 
                             $ \hat{x}-$direction when  the charge and the  thermal imbalance diffuse along the Hall bar.  In the presence of 
                             charge imbalance,  Eq.(\ref{vh})  is violated, so that the charge   
                             imbalance in the bulk has now be defined  as 
\bea
\rho_{I} =   n_\square \left ( c_{+,A} -  c_{+,B} -2\:  s_+\right ),
\label{imbo}
\enea  
 with reference to the remarks after  Eq.(\ref{conQ})\cite{nota}. 
 
 It follows that   constitutive  equation, Eq.(\ref{conQ}), can be rephrased  in the present scheme. Extracting from the difference of Eq.s(\ref{2e4bp}) 
 an expression for $ \delta {\bf J}_{n}\cdot \hat{x} $, which  appears on the left hand side  of Eq.(\ref{conQ}), we set 
                              \begin{widetext}

%   \bea
 % e \:\frac{ \partial }{\partial x} \left (  \delta {\bf J}_{n}\cdot \hat{x} \right ) \sim  \frac{ v_F }{{\cal{V}} } \: \left ( c_{+,A} -  c_{+,B} -2\:  s_+\right ) \equiv   \frac{ 1 }{ n_\square }\: \rho _I (x)  \: \frac{ v_F}{{\cal{V}} }. 
% \label{ncur}
  % \enea
 %   where  $   {\cal{V}} = [\ell w d]$  is the total   volume and $ \delta  J_n   $ is the particle current density of dimension $[ energy / t \ell ^2] $ along $x$. 
 
   \bea
  - 2 \partial_x \left \{  \frac{1}{2w} ( \sigma _1 -\sigma _2 ) \left [ (s_1-s_2)_A + (s_1-s_2)_B \right ] +  \frac{1}{2w} ( \sigma _1 +\sigma _2 ) \: 2\:
  s_+ \right \} =   - 2 \:  \frac{ v_F }{{\cal{V}} } \:   (  c_{+,A} -  c_{+,B} -2\:  s_+).
   \label{cona}
   \enea
        \end{widetext} 
  We remind that we assume for simplicity  that our  bulk 
           potentials satisfy  the mirror symmetry between opposite sides of the edges, $s_{+A} \approx - s_{+B } = s_+ $. This is a weak restriction 
           that could be lifted at the cost  of clarity, but does not invalidate the core of our arguments and our results.  The space derivative on the
           left hand side of  Eq.(\ref{cona}) defines the length scale for diffusion (in units of $L$),  when the  space dependence of the unknown 
           potentials $ c$'s and $s$'s has been determined. 
           %With $ \lambda_Q \sim 100 \: \mu m $ we  obtain  $ L \sim  1\: \mu m$ if we use a resistance $ R \sim 500\:  \Omega$.
  
   The  set of nine differential equations can be simplified by  neglecting the dependence on  $\Delta T$, which corresponds to perform 
   the derivation at   $\Delta T =0 $.  A  straightforward but boring derivation ( see Appendices) leads, to first order in the dimensionless 
   model parameter $\tilde {g }=   g \:  (2\gamma +g ) L^2  t_0 ^2 $, which quantifies the coupling between bulk and edge, to a 
   two-equation  set in  the unknowns  $\overline{T} $ and $\rho_I$.  Assuming that $-\tilde \sigma \: {\cal{E}}=  \rho \: u $, we  get:
        \begin{widetext} 
     \bea
   \kappa _{xx}\:( -\nabla ^2_\tx \overline{T} ) +   \alpha _{xy}  \nabla _\tx \left ( \overline{T} \:  {\cal{E}}L \right )= - 4 \: \left (\frac{
   \epsilon +{\cal{P}}}{n\: e} \right )_{3d} \:   \frac{ v_F }{ n_\square {\cal{V}}} \:\frac{   \sigma _1\sigma _2 }{(\sigma _1^2 +\sigma _2 ^2)} \: 
   \rho_I L^2. \hspace{5cm} \label{eqpx} \\
      \partial ^2_{\tilde{x}}  \rho_I +\frac{  2\: w \: L  \: v_F }{( \sigma _1 +\sigma _2 ) {\cal{V}} } \:   \partial _{\tilde{x}}
       \rho_I   - \tilde{g} \:  \rho _I  \hspace{10cm} \nonumber \\
 =- \frac{e\:n_\square}{L} \left (\frac{n}{ \epsilon +{\cal{P}}  }  \right )_{3d}   \frac{ \sigma _1 -\sigma _2 }{ \sigma _1 +\sigma _2 } \:
 \frac{w}{4} \:  \partial^2_{\tilde x} \left [\left \{\left (  \frac{1}{\sigma_1}-  \frac{1}{\sigma_2} \right ) \:  \kappa _{xx}  +\left ( 
 \frac{1}{\sigma_1}+  \frac{1}{\sigma_2} \right ) \: \alpha _{yx} \left (\frac{ \epsilon +{\cal{P}}}{n\: e} \right )_{3d} \right \} ( 
 -\partial_{\tilde x} \overline{T} )
 \right .\nonumber\\
 \left . +\left (  \frac{1}{\sigma_1}-  \frac{1}{\sigma_2} \right )  \:  \kappa _{xx} \: \left ( \frac{\alpha _{xy} }{ \kappa _{xx}}\: {\cal{E}}
 \: L   \right ) \: \overline{T}\right ]. \label{permu}
    \enea
       \end{widetext} 
       The system is written in terms of the  dimensionless space coordinate $ \tx = x/ L$.  $L $ is a  length unit at $B= 2 \: Tesla$ that will
       be found in Section V.  It is assumed to depend on  $B^{-1}$, as the classical cyclotron radius  $ r_c = p_F c / eB$.  The length scale of 
       variation of   $  \rho_I$  in Eq.(\ref{permu}) is $a^{-1} L$, where 
       \bea
        a \equiv \frac{  2\:  w L }{( \sigma_1+\sigma _2) {\cal{V}}  } \:  v_F ,
        \label{equa}
        \enea
    fixes the length scale for the particle/hole  imbalance  relaxation. Note that   Eq.(\ref{resa}) makes  the parameter $a$  independent of
    $d$.  $ \left(\epsilon + {\cal{P}}\right )_ {3d }$  is the enthalpy per unit volume. An important   parameter  is 
    the  effective carrier density $ n_{3d} \sim {\cal{N}}_{3d}/ {\cal{V}}$, whose determination we discuss 
     at length in Appendix C.  Here, we choose to  define the   thermal energy per particle (and unit charge) as:
     
\bea
 Q \equiv   \frac{\left ( \epsilon + {\cal{P}} \right )_{3d}  {\cal{V}} }{e  {\cal{N}}_{3d}}\equiv   \left ( \frac{ \epsilon + {\cal{P}}} {n\: e}\right )_{3d}. 
\label{ququ}
\enea
Choosing  $Q = 0.12 \times 10^{-3} V$ for the enthalpy  per unit particle, we are able to match reasonably well the values expected from a 
 Fermi Liquid approach and those experimentally found for $\alpha _{xy}$, $\kappa _{xx}$ and $ e_N$ (see Appendix C). 
  ($ n \equiv  n_{3d} \sim{\cal{N}}_{3d} / {\cal{V}} $ will have to be  self-consistently determined (see Eq.(\ref{denso}))).

      ${\cal{E}}$  appearing in Eq.(\ref{eqpx})  is the transverse applied electric field close to  $x =0$: it depends on $ \rho _I $ 
      away from $x=0$, and should be  self-consistency  recovered from  the transverse potential generated by $ \rho _I $ at any distance. 
      For the sake of simplicity, here  we linearize the corresponding equations, by considering two main $x-$regions: one close to the contacts where the external 
      current $J_y = \tilde{\sigma} {\cal{E}}_y $ is applied ("near zone")  and one "far" from $x=0$ ("far zone"), where the influence of 
      the externally applied field  vanishes. ${\cal{E}}\equiv {\cal{E}}_y $ is  the applied field at $x\approx 0$.  
      We keep  ${\cal{E}} \neq 0$  constant within each zone.   In the  far zone   ${\cal{E}}  ={\cal{E}}_{nl}$  is self-consistently determined,
      giving rise to the non local potential $v_{nl}$.   
      %In this way we get the  dependence of the transverse voltage on the magnetic field $B$  in the zone "far"  from $x =0$. 
             
%However, the system of Eq.(\ref{eqpx},\ref{permu}) is incomplete. There are other physical phenomena  that have been overlooked in deriving the system, which also require some discussion.  
 
 At $x=0$,   the boundary condition for $\partial _x\overline{T} ( x=0)$ is provided by the Nernst coefficient:
    \bea
 \left .     \partial _{x}  \overline{T} \right |_{x=0}  =  \frac{sign (B)}{ ( \sigma ^{-1}\alpha )_{xy} } \: {\cal{E}}_y(x=0).
 \label{bound}
 \enea  
 It will be shown in Section that this boundary condition together with Eq.(\ref{perpp}) implies $  \delta \mu_I  \sim  k_B \: T$  (see Eq.(\ref{mui})). 
 
 Our derivation has ignored the relaxation  of $\partial \overline{T}$  due to inelastic scattering processes including acoustic phonon excitation
 Away from the origin, the length scale  for these                relaxation processes  is $ \sim 10 \: \mu m $\cite{Orlita2008}.

 \subsection{Solution of Eq.s(\ref{eqpx},\ref{permu}) for $x \sim 0$}
\label{soleq}

To solve Eq.s(\ref{eqpx},\ref{permu}) for $x \sim 0$, we set   $\tx = x/L$ and introduce 
a dimensionless temperature $ \tau = (\overline{T} / ^o\! K) $, as well as a  
   dimensionless imbalance energy $\delta \mu_I = e \rho_I  L^2 / (k_B \: ^o\! K)$.
As a result, we can write the system of differential equations for 
$\tau$ and $\delta \mu_I$ as 
 
 \bea
 \partial  ^2_{\tilde x} \: \tau  -  \partial _{\tilde x}\: \left ( \theta   \: \tau \right )  =   \tilde{\beta}  \: \delta\mu_I , \:\:\:\:\: (a) \:\:\:\:\:
\nonumber \\
    \partial  ^2_{\tilde x} \: \delta\mu_I  + a\:  \partial  _{\tilde x} \: \delta\mu_I  
       - \tilde{g} \: \: \delta\mu_I   =\partial  ^2_{\tilde x} \: 
  \left \{ k \: \partial _{\tilde x} \: \tau - \theta \: u  \:\tau  \right \}. \:\:\:\:\: (b) \nonumber \\
   \label{tau1} 
  \enea
Eqs.(\ref{tau1}) depend on  the dimensionless parameters 
 \bea
  \hspace{.5cm}
  \theta  \equiv  sign(B) \frac{\alpha _{xy}  \:  {\cal{E}} \: L }{  \kappa_{xx} },\:\:\nonumber\\
    \tilde{\beta} \equiv \left (\frac{ \epsilon + {\cal{P}}  }{ n\: e^2}\right )_{3d}  \frac{   k_B  }{  \kappa_{xx} }\:
    \frac{ v _F }{ n_\square {\cal{V}} }\:  \frac{ 4 \:   \sigma _1\sigma _2 }{(\sigma _1^2 +\sigma _2 ^2)}  \; , \nonumber\\
  \label{parq}
 \enea
 as well as on the parameters $k$ and $u$, whose full expression is given in   Eq.s(\ref{ku},\ref{uu}),
 which depend on the imbalance ratio $q$, 
   \beq
 q=  \frac{   \sigma _1-\sigma _2 }{[\sigma _1^2 +\sigma _2 ^2]^{1/2} },
 \label{qbet}
 \eneq
that is linear in  the difference between the particle and the hole
conductivities.  The longitudinal conductivities $\sigma _i$'s are fairly isotropic and, 
therefore, they are expected to be pretty insensitive to the orientation of the magnetic field $B$. 
However,  we expect that the difference in chemical potential  between region A and region B  
changes sign when  $B$  is flipped,  as  flipping   $B$  implies exchanging   particles and  holes
with each other.  This effect, which we discuss in the following, corresponds to 
breaking the symmetry under $B \to - B$. 
 
For $x \sim 0$, it is  ${\cal{E}} \propto \theta$ that fully determines the  current in the $\hat y$ direction.  
Consistently with the experimental data, we assume   ${\cal{E}} \:w= 10^{-4} V$  takes the value given by the experiment. 
Moreover,  we also assume $B >0$ (the case $B < 0 $ will be analyzed in Section VI, when discussing the non local voltage). 
It is important to remark that changing the sign of $B$  corresponds to having  $q\to -q$, that is, to 
exchange with each other the Hall dynamics of particles and holes. 
 
 A more quantitative analysis will be given in the Appendices, by establishing the  numerical estimate of the parameters 
 involved. Here  we discuss the general features and our specific approach to the solution of the problem.  

 The  bath temperature of $1$ K corresponds to the value  in the experiment.  The longitudinal conductivities $\sigma_{1,2} $   
 are fixed by assuming the reference resistance $R \sim 10$~k$\Omega$ and the   inverse conductivities require an aspect ratio   
 $ 1/ \sigma _{3d}=   R \:\ell \: d / w $,  according to Eq.(\ref{resa}).  Inserting  $ \ell \sim 4\: w \sim  400 \: \mu m $, 
 $d \sim $~1~\AA, $L = 0.1\:  \mu m $ and the Fermi velocity $v_F=10^{8} \: cm/sec$ in the parameter $a$ defined in Eq.(\ref{equa}),
 gives  $ a \sim   10^{-3} $, so that the scale of variation for $\rho_I$ is $a^{-1}L \sim  100 \: \mu m$.  
The appropriate order of magnitude for the length scale $L$ is self-consistently determined in the next Section. 
%The choice   $L  \sim 1\: \mu m $  turns out to be consistent with the scale of variation of $J_n$ as shown in the Appendix.  

If the edge/bulk leakage parameter $ \tilde g \propto 10^{-6} $,  it can be shown (see  Eq.s(\ref{verosys}) in the Appendix D) 
that, employing the parameters  defined in Eq.(\ref{bet}), the system Eq.(\ref{tau1})  can be cast in a form in which all known 
quantities are  ${\cal{O}} (1)$, which is particularly 
amenable for   drawing plots from the numerical data.  
 Nevertheless,  for the general discussion here we will keep using  Eq.s(\ref{tau1}) as our 
 reference, as they appear to be more appealing for the sake of the physical interpretation
 of the results. 

 We now discuss the general features of the system by using an approximate analytical solution  in the region $ x\approx 0$ and 
 accounting for the boundary conditions of Eq.(\ref{bound}).

 % the length  $a^{-1} L \sim  10^3 L $ determines the space variation of the unknown functions $\delta \tilde{x} \sim 10^{-3} $.    an  order of magnitude  $10^{-6} $ factorizes in  Eq.(\ref{tau1}, (b)) and can be dropped. 

      We show in Appendix  E that, inserting    Eq.s(\ref{tau1},a)  for $\overline T$ into  Eq.s(\ref{tau1},b), one 
      recovers  a higher order  equation for $\delta \mu _I $. In dimensionless units, this  can be cast in the form of set of 
      first order differential 
      equations as:
      
      \bea
\left \{ \begin{array}{ll}      \partial  _{\tilde x} \:\delta \mu_I = \xi \\
\partial  _{\tilde x} \:\nu = \delta \mu _I \\   
 \partial  _{\tilde x} \: \xi  =  -( a  - \theta   -k  \tilde{\beta} ) \: \xi +[  ( a   - u  \tilde{\beta} ) \:   \theta +\tilde{g}   ] 
 \: \delta  \mu_I  -\tilde{g} \: \theta \:  \nu  , 
       \end{array} \right . 
       ,\nonumber
       \enea
       \bea
       \label{sys}
         \enea
with the additional equation for $\tau = \overline T/ \: ^o\! K$:
\bea
 \partial  ^2_{\tilde x} \: \tau  -  \partial _{\tilde x}\: \left ( \theta   \: \tau \right )  =   \tilde{\beta}  \: \delta \mu_I.
  \label{tau}
 \enea
 
  It is useful to shift $ a- u  \tilde{\beta} \to a $ and to define $ F =  ( a  - \theta  -\kappa  \tilde{\beta} )$ and $ G = 
  a \: \theta +  \tilde{g}   $,   where  $ \kappa = k-u $. The boundary conditions are satisfied by $F>0$ and $G,\theta, \tilde{g}  < 0$. 
  The system of Eq.(\ref{sys}) provides three possible solutions. The corresponding eigenvalues  characterize their  decay rate in real space 
when moving  away from the applied perturbation. One eigenvalue is real and the other two may be still real or  complex conjugate.  
  It is remarkable that in both cases  the eigenvalues  $ \lambda _{1,2} $ are independent of $\theta $, if and only if 
  $\kappa \propto q = 0 $.  Indeed, it is   shown in Appendix E that, for $\kappa = 0 $, the three eigenvalues  are given by:
  \bea
  \lambda _{1,2} = - \frac{a}{2} \pm  \sqrt{  \frac{a^2}{4}- |\tilde{g}| },\:\:\:\: \lambda _3 = -\theta \: sign[\tilde{g}].
  \enea 
% where $ \lambda _{1,2} $ do not depend on $\theta$.
  If $ \frac{a^2}{4}- |\tilde{g}| > 0$,  the three eigenvalues  are real, otherwise  $ \lambda_{1,2} $ are complex conjugate. 
  Let us consider the case $\kappa =q=0 $ for a while.  It follows that, for $\kappa =0$, only the solution corresponding to  
  $\lambda _3$  is relevant. It takes the form   
  \bea
\delta \mu_I= \eta   \theta^2 \: e^{ \theta x} 
\label{mumu}
\enea
 (with $\eta$ to be determined in the following).  Here, we assume  $\theta <0$ and $\tx >0 $. In any case, 
 we will always choose the decaying solution for  $\tx >0 $, which fixes  the sign of ${\cal{E}}$ for a given 
 orientation of $B$ orthogonal to the strip.  From Eq.(\ref{tau}), we obtain:   
 \bea
  \tau = 1 + \eta \: \tilde{\beta} \: \theta \: \tx \: e^{\theta \tx}.
  \label{croa}
   \enea
  Imposing  the boundary condition at $x=0$ as per Eq.(\ref{bound}), to ${\cal{O}}(\theta ^2 )$ we obtain: 
\bea
\left .  \partial _\tx \tau \right |_{\tx=0} =  \eta \: \tilde \beta \: \theta =
\frac{\kappa_{xx}}{ \: ^o\! K \alpha_{xy}( \sigma ^{-1}\alpha )_{xy} }\:  \theta 
 \label{taus}
 \enea
 which gives $  \tilde{\beta} \eta\sim 40$, once the parameters are chosen as   described in  Appendix C. 
 From the definition of $\tilde{\beta}$, the boundary condition Eq.(\ref{taus})  can be written as:
\bea
 \eta \: \frac{Q}{e}\:  \frac{   \:^o\! K \alpha _{xy}  ( \sigma ^{-1}\alpha )_{yx} }{\kappa_{xx}}\:  \frac{k_B}{ \kappa_{xx} }\:  
 \frac{2\:  v_F   }{n_\square {\cal{V}} } \:\frac{ 2 \: \sigma _1\sigma _2 }{(\sigma _1^2 +\sigma _2 ^2)}\:  \sim  1.\nonumber\\
 \label{vrig}
 \enea 
 This provides an estimate for $\eta \approx 2.35 \times 10^{11}$.
 
    Our definitions of the physical quantities are fully consistent, as we show in the following. 
   
% In the absence of charge imbalance the heat current  in the $\hat x$ direction is conserved. 

Let us  now   add  some  charge  imbalance  at  time $t=0$.  For  ${\cal{E}} \neq 0$, 
a thermal  flux  $-\partial _\tx  \delta J_Q =  -e \: Q \: \partial _\tx \delta J_n$ 
moves from $x=0$  into the bulk, parallel to the $\hat{x}$ axis.  From Eq.(\ref{jqsigma}), by differentiating 
Eq.(\ref{croa}) and using the definition of $\theta $  in Eq.(\ref{parq}), one obtains  
 \bea
- \left .  \partial _\tx   J_Q \right |_{x=0}  = \left [ \kappa_{xx}  \left . \partial _\tx^2 T  \right |_{x=0}  -\alpha _{xy} {\cal{E}} L \: 
\partial _\tx T \right ]/ 2L\nonumber\\
  =  \kappa_{xx} \:  \eta\tilde \beta\theta ^2 \times \: ^oK/2L = (\partial _\tx T ) \alpha_{xy} \:  {\cal{E}} /2 .  \hspace{1cm}
     \label{ch}
 \enea 
 (We have divided by $2$ because we only consider the flux in the direction $x >0$.) 
 $ \partial _\tx   J_Q $  is no longer vanishing as from Eq.(\ref{tau1} (a)) for $\mu_I =0$. Instead,   it 
 now generates a thermal gradient.
 
  On the other hand, 
 by substituting  $v_F /( n_\square {\cal{V}}) $  from  Eq.(\ref{conQ}) in Eq.(\ref{vrig}), we get:
 \bea
- e\:Q\:   \partial _x  \delta J_n \:L^2\:  \frac{  \alpha _{xy}  ( \sigma ^{-1}\alpha )_{yx} }{\kappa_{xx}}\:  
\frac{2}{ \kappa_{xx} }\: \frac{ 2 \:   \sigma _1\sigma _2 }{(\sigma _1^2 +\sigma _2 ^2)}\: = \theta^2 , \nonumber\\
\label{tem}
\enea
 where we have identified  the fluctuation of the chemical potential,  $ \rho_I L^2 /( k_B  \: ^oK) \equiv  \delta  \mu_I  =  
 \eta \theta^2$, on the r.h.s.  from  Eq.(\ref{croa}).  Using  Eq.(\ref{ch}), Eq.(\ref{tem}) reads:
\bea
 L \:  \partial _x T \:  \frac{  \alpha _{xy}  ( \sigma ^{-1}\alpha )_{yx} }{\kappa_{xx}}\: \: 
 \frac{ 2 \:   \sigma _1\sigma _2 }{(\sigma _1^2 +\sigma _2 ^2)}\: \approx  \theta  
 \label{devt}
\enea
which gives $  \partial _\tx T  \sim 40\:  \theta  \: ^o\! K  $ once more,  as in Eq.(\ref{taus}) because the 
ratio $2 \:   \sigma _1\sigma _2 /(\sigma _1^2 +\sigma _2 ^2) \approx 1 $. 
%By putting $   \partial _\tx T \sim  q \eta \: \tilde{\beta}\theta $,  we get $\eta \: \tilde{\beta}  \sim  40 $ again.  $ \tilde{\beta} $ is found to be $ \tilde{\beta} \sim 10^{-9}$, so that $ \delta \mu _I \sim 10^{-1}  q$. The fact that this heuristic argument leads to the same conclusion as before, confirms the  reliability of the definitions  introduced in  our model.

\section {Ettingshausen parameter}
\label{sec5}
\setcounter{equation}{0}

The first step is to estimate the leakage factor $ \tilde g$ and, consequently, the actual number of
free carriers,  $ {\cal{N}}_{3d}$, that have been redistributed between the areas $A$ and $B$ of the Hall bar. 
In dimensionless units, the energy associated to the transverse  voltage across the Hall bar (in the ${\hat y}-$ direction) 
is given by: 
% Following the derivation of the Appendix C, Eq.s(\ref{verosys}), 
\bea
 v_{y}  = e\:  ( c_{+,A} -c_{+,B}) /( k_B  \:^oK). 
 \label{vol}
 \enea
%Eq.(\ref{vol}) defines ${\cal{N}}_{3d} $. 
Following the derivation of  Appendix F, we get,  to ${\cal{O}}[q^2]$ and  ${\cal{O}}[ \tilde{g}]$ :  
% [from hgteNeo4Summa, where  where  the substitution $ e/ d \to ( \epsilon +{\cal{P}} ) / (  n_\square e) $ is required],
 
\begin{widetext}

 \bea
   \partial ^2_{\tilde x } \:  v_{y}    - \tilde g \: v_{y} 
   =   \tilde g  \:  \frac{e}{L} \:  \left (  \frac{1}{\sigma_1}+ \frac{1}{\sigma_2} \right ) \: 
   \frac{w}{4\:  k_B Q}\: \left [ \alpha _{xy}   \: {\cal{E}} L \: \frac{T}{^o\! K}-  \left ( \kappa _{xx}  -
   \frac{ \sigma _1-\sigma _2}{\sigma _1+\sigma _2}\: \alpha _{xy}\: Q \right ) \: \partial _{\tilde x} \tau  \right ]. 
   \label{hallvolt1}
 \enea
\end{widetext}
In order to establish the consistency, in this Section we  concentrate   on  
the area around $x\approx 0$ where the external bias is applied.  We will fix $ {\cal{N}}_{3d}$, 
together with the scale $L$, by assuming an applied electric field at the origin ${\cal{E}} \sim  V/m$.

The r.h.s. includes a term $ \propto  [ \alpha _{xy}   \: {\cal{E}} L \: {T}/{^o\! K} -  \kappa _{xx} \: \partial _
{\tilde x} \tau ] $ which describes the particle flux $ J_n$  flowing  away from   $x\approx 0$. 
As metallic contacts are applied  at $x\approx 0$, we   expect a flow in the contacts of the 
charge carriers lost because  the first two 
terms.  The last  term $\propto  q\:  \partial _{\tilde x} \tau$  relates  the 
transverse voltage to the  thermal gradient in the presence of the orthogonal magnetic field. A thermal gradient in 
presence of a current with a magnetic field is named Ettingshausen effect. We now focus on it as,   in our case,  
it is proportional to the charge imbalance.  If we assume that, close to the origin, the external source sustains the 
particle flux  $  \partial _{\tilde x} J_n$ given by Eq.(\ref{ch}) into the contacts, so that 
$  \partial _{\tilde x}\left [ \partial _{\tilde x} \tau - \theta  \: \tau \right ] \approx 0 $, according to 
the definition of $\theta $ given by Eq.(\ref{parq}),  an estimate of the Ettingshausen parameter can be obtained. 
Self-consistency also allows to determine the effective  $3d$  carrier density $ {\cal{N}}_{3d}$, as we show in the 
following.

For $x \sim 0$, we can neglect the term $  \partial ^2_{\tilde x } \:  v_{y}$ at the l.h.s of Eq.(\ref{hallvolt1}). 
Within this approximation,  $\tilde g$  drops out for $x \sim 0$, where the charge distribution is 
fixed by the applied  electric field. Of course,  $\tilde g$  will play an important role in the next Section, when estimating 
the non local transverse voltage. By keeping just the last term in Eq.(\ref{hallvolt1}), which we denote   by $v_\perp$, we get
\bea 
  v_{\perp}  \approx  \frac{e}{L} \: q  \left (  \frac{1}{\sigma_1}+ \frac{1}{\sigma_2} \right ) \:  \frac{w}{4\:    k_B }\: 
\alpha _{xy}\: \partial _{\tilde x} \tau .
\label{perpp}
\enea
It is important to realize that, for $x \sim 0$, where the current is fed in, 
$v_\perp$ does not  flip its sign when $B $ changes sign. This is because $ \alpha _{xy}$ is an  odd function of $B$ and, 
when $B$ changes sign, particles and holes exchange their position between area A and B,  so that  $q$ changes sign, as well.   
 Eq.(\ref{perpp}) can be usefully rewritten, according to the definition of $\tilde\beta $ given by  Eq.(\ref{parq}), as   
\bea 
  v_{\perp}  \approx \tilde{\beta}^{-1}\frac{ a \: q }{n_\square L^2 } \: Q\: \frac{(\sigma _1+\sigma _2)^2}{2(\sigma_1^2+\sigma_2^2)}
\frac{\alpha _{xy}}{\kappa _{xx} }\: \partial _{\tilde x} \tau .
\label{perp1}
\enea
where $a$ is defined in Eq.(\ref{equa}).  To set up the selfconsistency, we insert  
$ v_{\perp} = e \: q {\cal{E}} w /( k_B \:  ^o\! K)  $, $  {\cal{E}} L =( \kappa _{xx} /\alpha _{xy}) \theta $ in 
Eq.(\ref{perp1}). From Eq.(\ref{devt}), observing that $\frac{\sigma_1^2+\sigma_2^2}{2 \sigma _1\sigma _2} \sim 1$, we get:
\bea
 \frac{\kappa _{xx} }{\alpha _{xy}} \: \theta \:  \frac{ 1 }{ k_B}  \:  \frac{ w }{ L}\approx \tilde{\beta}^{-1} \:\frac{a   }{n_\square L^2 } 
 \frac{Q}{ e}\: \frac{\theta}{ (\sigma ^{-1} \alpha ) _{xy} },
    \label{consist0}
\enea
consistent  with the definition of $\tilde{\beta}$.

 With ${\cal{E}} \sim V/m$ we get 
\bea
\theta = \frac{\alpha _{xy}}{ \kappa _{xx} } \:  {\cal{E}} L   \sim 10^{-5}
\enea 
and 
  \beq
      \delta \mu_I \sim   \eta \: \theta ^2   \sim 2.3 \: .
  \label{mui}
 \eneq   
  Comparison with Eq.(\ref{imbo}) requires that 
\bea
 \frac{ \rho _I}{e} = n_\square  \frac{\delta  \mu _I}{e}  \frac{k_B}{e} \: ^o\! K=  2.3 \times \frac{10^{10}}{cm^2} 
 \frac{86.17\times 10^{-6}}{27.16 \times 0.529 \: \AA}\nonumber\\
  =   14 \times 10^4 \left ({cm^2 \AA}\right)^{-1}.  \hspace{1cm}
  \label{cimbo}
 \enea
 This is fully   consistent with  the initial definition $ \delta \mu _I = e \: \rho _I L^2 / k_B  \: ^o\! K$, provided 
 we set $L \sim 0.1 \: \mu m$:
 \bea
   1=  n_\square \: L^2
   = \frac{10^{10}}{cm^2} (0.1\:  \mu m )^2. 
   \label{valo}
    \enea 
          Here further consistency requires that 
    \bea 
     \frac{ \rho _I}{e} =  14 \times 10^{12}  cm^{-3}  =q\:  n_{3d}.
     \label{denso}
      \enea
      With a density $ n_{3d}    = 1.3 \times 10^{15} \: cm^{-3} $, at $B = 2 \: Tesla$, the bulk Hall conductance  is
      \bea
       \sigma _B =  n_{3d} \: \frac{ ec}{B} = 10^{12} sec^{-1}, 
       \label{sigB}
       \enea
        a value  roughly consistent with  the ratio $   \kappa _{xx}^{FL} /{  \alpha _{xy}^{FL}} $ 
        given in Eq.(\ref{ratio}), which provides $\sigma _B \approx 5 \times 10^{12} \: sec^{-1} $.  
        According to Eq.(\ref{denso}), this density requires $ q \sim 10^{-2}$.   The inequality  $ \sigma _B / \sigma _{3d}  <<1$ is therefore confirmed.  

      The electrostatic  energy per unit volume  due to the  charge fluctuation is:
 \bea
 \frac{e}{2}  \: n_{3d} \: \rho_I L^2  = \delta \mu_I   \:  n_{3d}  =\frac{1}{2}\eta \theta ^2  \:  n_{3d} \approx  46.7  \frac{m\: eV}{ cm^2 \AA}.\nonumber\\
 \enea  
        It is remarkable that Eq.(\ref{perpp}) entails   the definition  of the Ettingshausen parameter, as we show  
        in the following, by resorting to the Fermi Liquid forms of the transport parameters 
        $\alpha _{xy}^{FL} $ and   $\kappa_{xx}^{FL} $.  The Ettingshausen ratio is:
     \bea
        P_{E} = \frac{ \partial _{ x} \bar{T}}{ |B|  J_y} 
        \enea
      
Now,  substituting the parameters given in Eq.(\ref{seit2},\ref{seit}) into  Eq.(\ref{perp1}), 
with 
\bea
   \frac{\alpha _{xy}^{FL}}{\kappa _{xx} ^{FL}}  = \frac{\sigma _B}{\sigma _{3d} }\frac{1}{Q} \frac{1}{
\left [1- \left ( \frac{\sigma _B}{\sigma_{3d}}\right )^2 \right ]},\nonumber
\enea
we get: 
\bea
v_{\perp}   = \tilde{\beta} ^{-1}   \frac{ q\: a }{ n_\square  L^2}  \frac{(\sigma _1+ \sigma _2)^2}{2(\sigma _1^2+ \sigma _2^2)} 
\frac{\sigma _B}{\sigma _{3d} } \frac{1}{
\left [1- \left ( \frac{\sigma _B}{\sigma_{3d}}\right )^2 \right ]} \: \partial _{\tilde x} \tau \nonumber
\enea
Posing $ e \: q {\cal{E}} w / k_B \: ^o\!K = v_{\perp} $  again, and dropping  
\bea
\frac{2\:(\sigma _1^2+ \sigma _2^2)}{(\sigma _1+ \sigma _2)^2}\left [1- \left ( \frac{\sigma _B}{\sigma_{3d}}\right )^2 \right ] \approx 1 ,\nonumber
 \enea
 we are left with 
\bea
\frac{e}{k_B } \: n_\square  L^2\:   \tilde{\beta} a ^{-1}  \: \frac{w}{L} \:  = \sigma _B 
\:\frac{  \partial _{ x} \tau \: ^o\!K }{\sigma _{3d}{\cal{E}}},
\enea
or
\bea
 P_E^{FL} = \frac{1}{B} 
\:\frac{  \partial _{ x} \tau \: ^o\!K }{\sigma _{3d}{\cal{E}}} = \frac{1}{k_B \: c} \:   
\tilde{\beta} a  ^{-1}  n_\square  L^2\: \frac{ 1}{n_{3d} } \:  \frac{w}{L}  .
\enea
With  $B = 1 \: Tesla$ and $  \tilde{\beta} a  ^{-1} \sim  10^{-6} $,
\bea
P_E^{FL} =1.1  \times10^{-14} \frac{cm^3 }{eV \: c} \: ^o\!K  \nonumber\\
= 1.7 \times 10^{-2}   \frac{ meter \: ^o\!K}{Amp\:  Tesla },
\label{ettinga}
\enea 
as $  Amp /meter = 8.97 \times 10^{9}  V/sec$.

   This is our first result, to be compared with  the case of Bismuth: $P_{E \: Bismuth}   = 7.5 \times 10^{-4}   \frac{ meter \: ^o\!K}{Amp\:  Tesla }.$ 
   At the bottom of Section VI we argue that consistency with  the measured NLR magnitude
   ( $ R_{NL} \sim k\Omega $) suggests that the actual  value of $P_{E}$ far from the source should be lowered of about 
   two orders of magnitude with respect to the one  reported in Eq.(\ref{ettinga}). 
   Given the uncertainities  in the effective bulk density, our result cannot be sharper.

\vspace{0.3cm}

\section{The transverse non local voltage}
\label{sec6}
\setcounter{equation}{0}

We now turn back to Eq.(\ref{hallvolt1}) and examine it for   $ x \sim D$, far away from the origin. 
Defining 
 \bea
  r =  \frac{ sign(B)\:  e}{L}\: \left (  \frac{1}{\sigma_1}+ \frac{1}{\sigma_2} \right ) \:  \frac{w}{4\:  k_B Q}\: \kappa _{xx},
 \enea 
  Eq.(\ref{hallvolt1})   reads:
  \begin{widetext}
  \bea
   \partial ^2_{ \tx } \:  v_{y}    - \tilde g \: v_{y} 
   =   \tilde g\: r\:  \left [  \frac{ \alpha _{xy} }{\kappa_{xx}}\: {\cal{E}}_{nl} L \: \frac{T}{^oK} - 
   \left ( 1 - \frac{ \sigma _1-\sigma _2}{\sigma _1+\sigma _2}\: \frac{\alpha _{xy}}{\kappa_{xx}}\: Q \right ) \:
   \partial _{\tx} \tau  \right ] \: ,  
   \label{Ahallvol}
 \enea
 \end{widetext}
where  $T =1 \:^oK $. 
For $x \sim D$, there is no applied electric field. However, as we explain in the Introduction, 
the thermal gradient  at the origin determines  a  charge imbalance that propagates in the 
$\hat{x}$ direction with a very low  relaxation rate. Once transported at $x \sim D$,  
the charge imbalance  generates the electric field $ {\cal{E}}_{nl} $, 
which is the source of the $B$-dependent  contribution to the  non local resistance. Note that the
product $ \alpha _{xy}\: {\cal{E}}_{nl} $ does not depend on $sign(B)$, although both  factors 
 do.  A consistency  condition can be set in Eq.(\ref{Ahallvol}) by posing  $ v_y = e {\cal{E}}_{nl} w/(2 k_B ^oK)$ 
(the factor $1/2$ arises from the observation that the electric field is not given by an external source but created when 
the charge difference accumulates close to  the edges).  Switching  to the variables  $ \tilde{\tx} = x/ a^{-1}L$   and $\tau '  =
a \tau $,  we get:

\begin{widetext}
 \bea
   \partial ^2_{\tilde{ \tx} } \:  v_{y}    +\left \{ \tilde g \: a^{-2}  \left [2\:  r \:  \frac{ \alpha _{xy} }{\kappa_{xx}}\:
   \frac{L}{w} \frac{k_B T }{e}  -1\right ]  - \frac{1}{\tau_Q^2}\right \}\:  v_{y} 
   =  - \tilde g\:a^{-2} \:  r \:  \left (1 - q\: \frac{\alpha _{xy}}{\kappa_{xx}}\: Q   \right ) \: \partial _{\tilde{\tx}} \tau ' . 
   \label{Ahallvolt6}
 \enea
   \end{widetext}
   In Eq.(\ref{Ahallvolt6}), we have added an extra term  $- { v_{y} }/{\tau_Q^2} $, to account for the intrinsic  relaxation.   
   In any case, even in the absence of this term,   $ v_{y}$ is a decaying function  at large values of $x$,  where the  thermal gradient 
goes to 0, as a consequence of the fact that $ \tilde g \: a^{-2} <0 $ and that  the quantity within the square brackets is positive.  
The l.h.s. of Eq.(\ref{Ahallvolt6})   implies a decay length $\lambda$ of the non local voltage. For $\tau _Q\to \infty$, one obtains   
 \bea
  \lambda  \sim  \left [  \frac{\sigma_1+\sigma_2}{2 \: \sigma_1 \sigma_2} \frac{ T \alpha _{xy} }{Q}-1  \right ]^{-1/2} \:  L/\sqrt{ \tilde g} .
  \label{length}
  \enea
  We argue that this length is well defined, because the difference in the square bracket is, on very general grounds,  greater than zero.
  Recalling that the Nernst coefficient $e_N = \sigma _{3d}^{-1} \alpha _{xy} \sim {\cal{E}}_y / \partial_x T $ and that 
  the orbital motion of carriers driven by the magnetic field is roughly circular, we recognize a torque acting on one carrier
  which generates a work  $  {\cal{L}}=  \delta x \: e {\cal{E}}_y $  per unit angle. On the other hand, in the same wedge  
  the energy per particle  provides a  thermal contribution to the change in the free energy    $ \sim s\:  \delta T  $, where $s =e\:Q/T$
  is the entropy per unit particle.  Ultimately,  the decay rate is related to a change of the free energy  per particle $ \delta f $ 
  where $f= u- T\: s $ is the drift due to the electric field. In fact,  $ \delta f =  -   {\cal{L}} - s \: \delta T $. As   $ \delta f $ 
  has to be negative in order for  the system to  evolve towards equilibrium, one obtains $ {\cal{L}} + s \delta T  >0$, with $ {\cal{L}} < 0$ 
 (as it is  performed by the external source) and $s \delta T > 0$ as  $\delta T  0$  is  induced by the source. 
  
  Indeed, one obtains  
    \bea
   2 r\: \frac{\alpha _{xy}}{\kappa_{xx}}\: \frac{L}{w}\frac{k_B T}{ sign(B)\: e} =   \frac{\sigma_1+\sigma_2}{2\: \sigma_1 \sigma_2}
   \frac{ T \left | \alpha _{xy}\right | }{Q}  > 1
    \enea
  for $ T = 1\:  ^o\! K$ and $ \sigma _1 = 7.19 \times 10^{14}, \:   \sigma _2 = 4.79 \times 10^{14}$, while, on the r.h.s. of  Eq.(\ref{Ahallvolt6}),  
  $ Q \: \alpha _{xy} /{\kappa_{xx}} \approx  0.12 \times 10^{-3} $.

It is important to note that the value chosen for the energy per particle, $e\: Q $, is $\sim k_B T$, but, in presence of
magnetic field, the magnetization energy $ -M\: B$ 
has to be added to it, which makes the argument of the square root in Eq.(\ref{length}) even more positive, when  $B $ increases.  

  It is also remarkable that the closer $ \sigma_1$ is to $\sigma_2 $, the longer the decay length $\lambda $ is.

   The space scale is $ a^{-1}  L \sim 100 \:\mu  m $ and, in this scale,  $ \lambda \sim 1$, or even larger. Indeed in our case the enthalpy
   per particle, including the magnetization work, as well,  is given by $ e Q =\left ( 0.12 \:10^{-3} - 4\times 10^{-4 }  |b| \right ) eV$, 
   where $b$ is the magnetic field in units of $2\: Tesla$.
   
  The inhomogenous term in Eq.(\ref{Ahallvolt6})  dominates  at distances $D \sim 3 \: a^{-1} L$ from the origin. 
   In the plots, we have chosen a phenomenological $B$ dependence $\tau_Q \propto B^{-1/2} $ and we have considered a Hall bar which is
   infinitely long   in the $x>0$ direction. 
   
    In deriving  Eq.(\ref{Ahallvolt6}),  $B$ was assumed positive. For negative  $B$ values,  $\alpha_{xy} \propto 1/B$ changes sign,
    while $ \kappa _{xx} $ is, in general,  an even function of $B$ and, in particular, here it is taken independent of $B$. No other  
    functional dependence on $B$ is introduced,
    except for $\tau _Q  \propto 1/ \sqrt{|B|}$.  However convergency of the solution of Eq.(\ref{mumu}) requires that  
    $\theta$ of Eq.(\ref{parq}) is an even function of  $ B $.  The prefactor $r$  as well as the imbalance $q$, and $\sigma_B$ of 
    Eq.(\ref{sigB})  are  odd in $B$, because  particle and hole exchange their role by flipping the magnetic field.  
    In solving the differential system,  the initial condition  for the integral of the imbalance chemical potential, 
    $\nu$, is also  odd: $\nu (0) = sign(B)$.

\begin{figure}
\begin{center}
\includegraphics[width= 0.9 \linewidth]{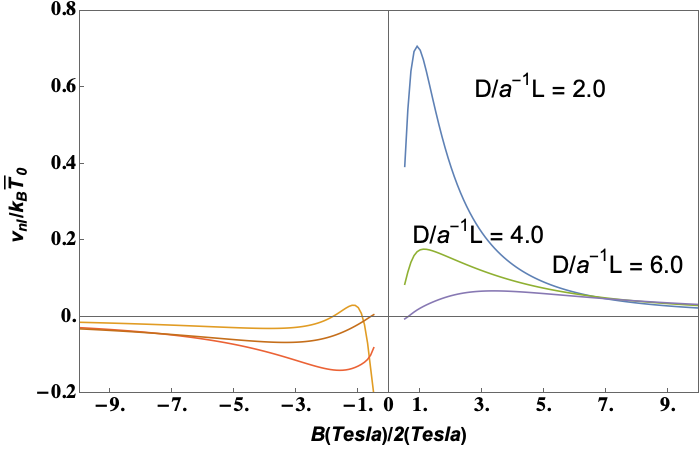}
\caption{Normalized  $[e \times$  non local voltage] vs $ b =B/(2\:Tesla)  $, at various distances from the origin: $D /a^{-1}L =2.,4.,6.,8$.  }
\label{vconb}
\end{center}
\end{figure}

   Fig.(\ref{vconb}) displays the main result of this work. We describe here qualitatively the picture that emerges from the model with
   the help of the plots of the relevant quantities. 
   
\begin{figure}
\begin{center}
\includegraphics[width= 0.9 \linewidth]{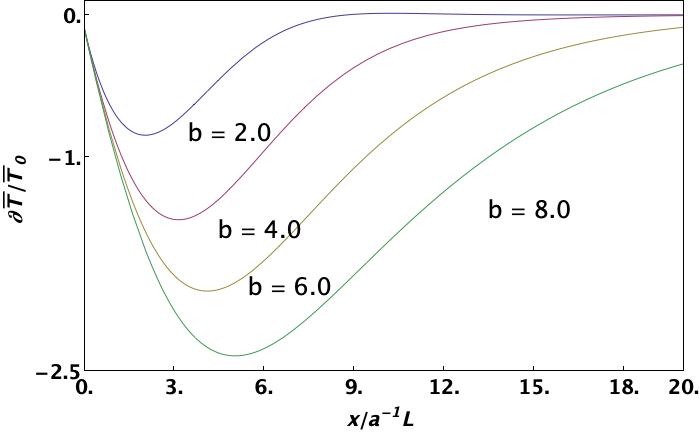}
\caption{Normalized  space derivative of the temperature vs distance $D /a^{-1}L $ from the origin, at  different magnetic fields ($ b =B/(2\:Tesla) =2.,4.,6.,8.$). }
\label{dTconxbpos}
\end{center}
\end{figure}

\begin{figure}
\begin{center}
\includegraphics[width= 0.9 \linewidth]{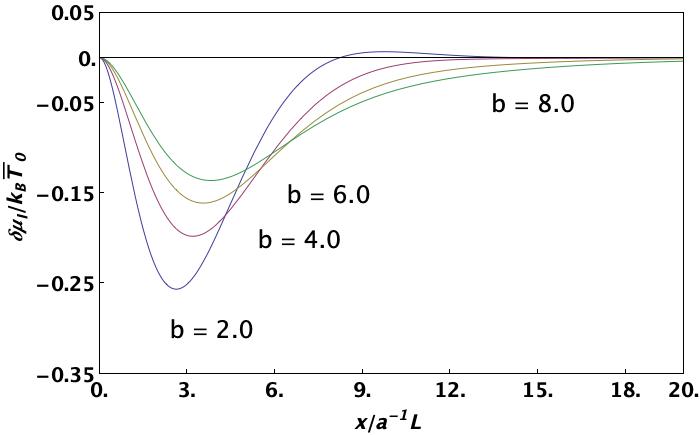}
\caption{ Normalized imbalance chemical potential  $\delta \mu _I $ vs  distance $D /a^{-1}L $ from the origin,  at  
different magnetic fields ( $b =B/(2\:Tesla) =2.,4.,6.,8.$) }
\label{muconxbpos}
\end{center}
\end{figure}

\begin{figure}
\begin{center}
\includegraphics[width= 0.9 \linewidth]{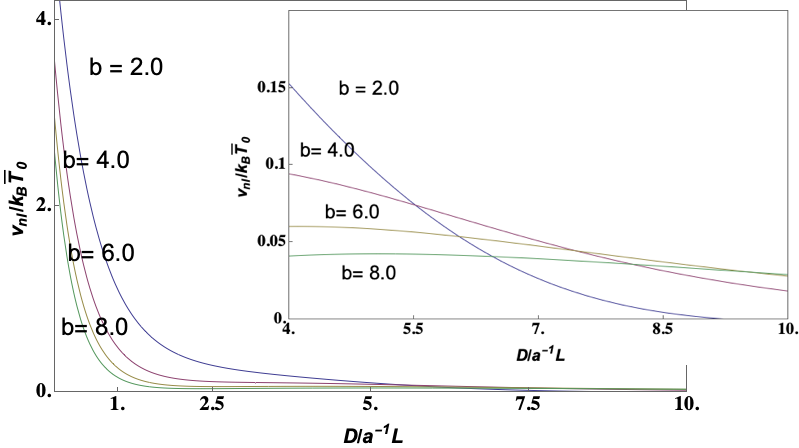}
\caption{Normalized  [$ e \times$  non local voltage] vs  distance from the origin $D /a^{-1}L$  at  different magnetic fields ($b =B/(2\:Tesla) =2.,4.,6.,8.$). 
Inset:  zooming of $[e \times$  non local voltage] at larger distances from the origin,  at  different magnetic fields ($b =B/(2\:Tesla) =2.,4.,6.,8.$)  }
\label{vconxbpos}
\end{center}
\end{figure}

     According to the sketch   in Fig.(\ref{scheme3}), the applied electric field ${\cal{E}}_y$ and the corresponding current $J_y$ 
     are oriented from  edge B (the lower edge of the picture) to  edge A (the  top edge of the picture), independently of the orientation 
     of the magnetic field $B$. When  $B> 0 $, the  carriers leaking in the bulk  from the edges  moving toward $x>0$ are particles close to edge A  
     and holes close to edge B.  The carriers moving in the other (opposite) direction, impinging on  the Hall bar boundary, have opposite sign, 
     but are assumed to be absorbed by the boundary and, so, they do not enter our discussion. For $x \sim 0$,  the system is assumed to be thermalized by the
     boundary, but an increase of temperature  with respect to the thermal bath is expected, due to the  Joule heat accompanying the applied current. 
     Let us consider the case  $B>0$ first. Nernst effect provides a thermal gradient which moves the carriers away from the applied field region.  
     The  increase  of  temperature drops relatively fast  away from the origin. This is reported in Fig.(\ref{dTconxbpos}) showing the thermal gradient.
     At large distances the temperature decreases at a rate decreasing with the distance, till 
     it becomes  constant. Under the effect of the thermal gradient,
     the carriers diffuse in the Hall bar as proved by the space dependence of the  difference in chemical potential between edge A 
     and edge B, $\delta \mu _I$,  which is plotted in Fig.(\ref{muconxbpos}) at increasing magnetic fields.  As the relaxation 
     time across the bar is rather  long, the carriers  can reach regions of the Hall bar where the effect of the applied field has vanished,
     so that  $\delta \mu _I$ keeps finite also at distances $D \approx 6 \: a^{-1}L$. This implies that a voltage difference develops between
     edges which is at even with the applied voltage.  For $B>0$, this gives a positive  non local voltage, which is reported in Fig.(\ref{vconxbpos}).
     
\begin{figure}
\begin{center}
\includegraphics[width= 0.9 \linewidth]{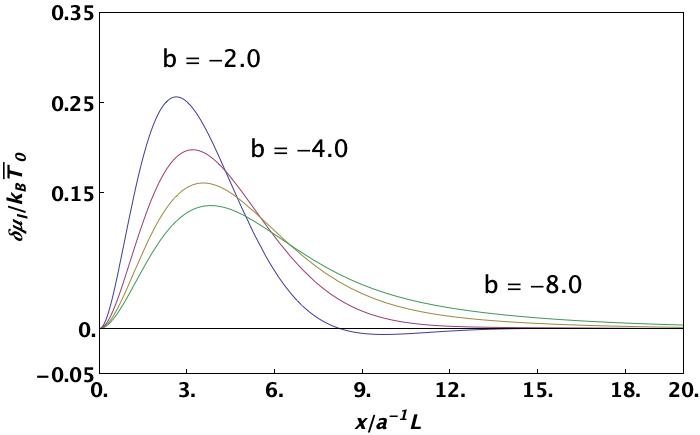}
\caption{  Normalized  imbalance chemical potential  $\delta \mu _I $ vs  distance $D /a^{-1}L $ from the origin, 
at  different magnetic fields ($ b =B/(2\:Tesla) =-2.,-4.,-6.,-8.$) }
\label{muconxbneg}
\end{center}
\end{figure}

\begin{figure}
\begin{center}
\includegraphics[width= 0.9 \linewidth]{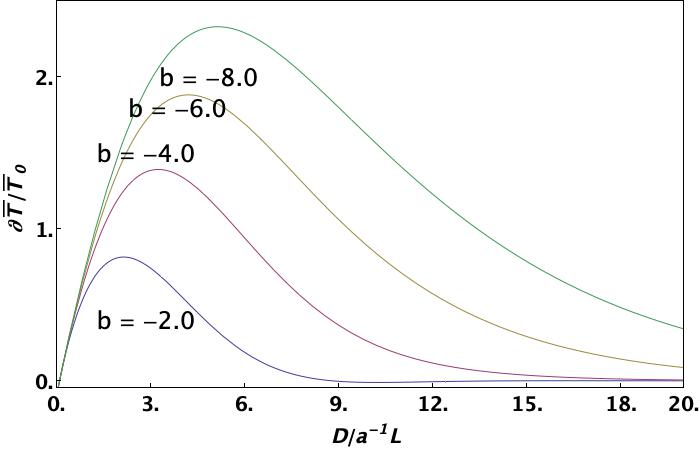}
\caption{Normalized  space derivative of the temperature vs distance $D /a^{-1}L $ from the origin, at 
different magnetic fields ($ b =B/(2\:Tesla) =-2.,-4.,-6.,-8.$). }
\label{dTconxbneg}
\end{center}
\end{figure}

      Let us now assume $B<0 $.  In this case the positive and negative carriers leaked 
      in the bulk of the Hall bar and drifting  at $x>0$ have opposite sign with respect to sketch $i)$, so that a positive $\delta \mu _I$ 
      develops between  the two boundaries (see Fig.(\ref{muconxbneg})) and the corresponding voltage difference is at odd with respect to 
      the applied one. This induces  heat diffusion away from the injected current and the thermal gradient is opposite to the one of   
      Fig.(\ref{dTconxbpos}), as plotted in  Fig.(\ref{dTconxbneg}).  This fact increases the distances at which the perturbation diffuses. 
      The result is that not  only the sign of the voltage correction  is opposite (see  Fig.(\ref{vconb}) for negative magnetic fields),
      but there is a marked difference  in amplitude between the contributions coming from opposite orientations of the magnetic field .   
  
\begin{figure}
\begin{center}
\includegraphics[width= 0.9 \linewidth]{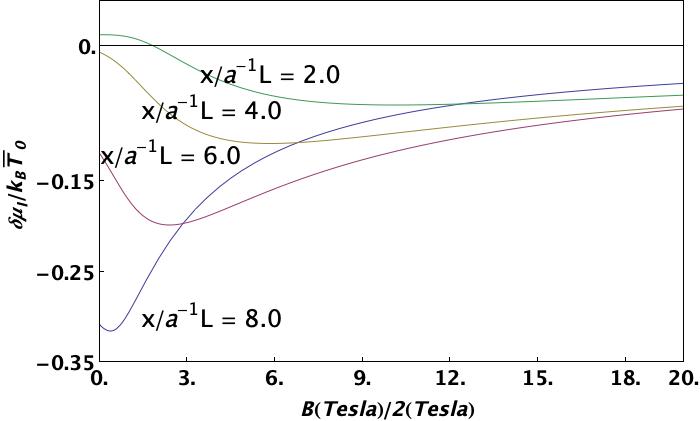}
\caption{  Normalized  imbalance chemical potential  $\delta \mu _I $ vs  $b =B/(2\:Tesla)$  at various  distances   
$D /a^{-1}L  =2.,4.,6.,8.$ from the origin }
\label{muconbpos}
\end{center}
\end{figure}

\begin{figure}
\begin{center}
\includegraphics[width= 0.9 \linewidth]{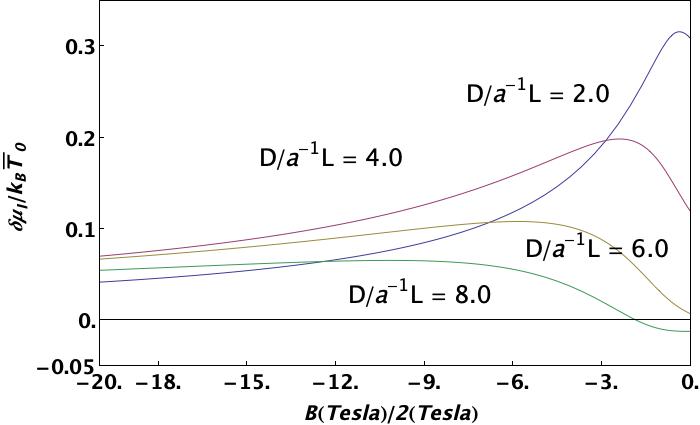}
\caption{  Normalized  imbalance chemical potential  $\delta \mu _I $ vs  $b =B/(2\:Tesla)$  at various  
distances   $D /a^{-1}L  =2.,4.,6.,8.$ from the origin }
\label{muconbneg}
\end{center}
\end{figure}

    While the non local voltage has an exponential  decrease  not far from the origin  (see Fig.(\ref{vconxbpos})), it is power law far
    from the origin  as shown in   the inset of  Fig.(\ref{vconxbpos}). The decay  toward  zero of the imbalance chemical potential  
    for larger magnetic fields  is slower the larger the magnetic field is, as shown in  Fig.(\ref{muconbpos}) and in  Fig.(\ref{muconbneg}).
    They are opposite in sign, but  the first one adds up to the applied voltage, the second one is subtracted.    
    It is remarkable that  the relative damping ratio with distance is clearly weaker for $B<0$. 
     
     The contribution to the non local voltage of Fig.(\ref{vconb}), appropriately scaled,  according to the experiment, by fixing the 
     constant energy scale $ k_B \overline{T}_0$, adds up  to  the non local voltage difference  derived in Ref.[\onlinecite{Nachawaty2018}]. 
     The corresponding transverse resistance  arises from   edge states only, which, along their path,
     suffer  some dissipation at the various contacts of the Hall bar.  While the contribution coming from the edges is fully symmetric with 
     reversing  the orientation of the magnetic field, 
     the contribution derived here introduces an asymmetry which is found in the experiment. The asymmetry tends to reduce with increasing $D$, 
       what is fully consistent with our plots.


    The differential system is fully homogeneous. Unknown functions depend on an overall scale normalization. 
    Close to the origin, the magnitude of  $\partial _x \tau $ is determined by ${\cal{E}}$, 
    field according to the Nernst boundary condition of  Eq.(\ref{bound}).  This is the reason why we could extract the 
    Ettingshausen parameter in the near zone. To extract the magnitude of the NLV  away from the origin, the differential 
    equation system should have been solved allowing for a selfconsistent   ${\cal{E}}$ field at any distance ($ \theta \to \theta (x)$) . 
    This has not been done. Therefore,   in the far zone the  magnitude of the NLV is strictly speaking undetermined. 
    We could try to use our  estimate of the  Ettingshausen parameter of Eq.(\ref{ettinga}) to determine the normalization, 
    by using $nAmp$ as a probe current  and $1\: Tesla$ as the reference magnetic  field, and extracting  a temperature 
    derivative $\partial T/\partial (a^{-1}L) \approx 10^{-2} \:^o\! K / ( a^{-1} L) $. However, when we plug this in 
    Eq.(\ref{Ahallvolt6}) and we evaluate the NLR by means of the same current,  we find that our estimate is two-three orders   
    of magnitude larger than the measured one, amounting to   $k\Omega$s.

\section{Summary and Final Remarks}
\label{sec7}
\setcounter{equation}{0}

Near the Dirac point, graphene Hall bars in an applied orthogonal magnetic field of the order of a few Tesla, exhibit 
 large nonlocal resistances which cannot be fully explained within the framework of the ``ideal''   Quantum Hall Effect (QHE) 
 in graphene. Here, we discuss  nonlocal resistances in a Hall bar configuration corresponding to an open circuit in the longitudinal direction,
 parallel to the length of the  Hall bar. In these conditions, one would   expect zero voltage between contacts in the transverse direction, 
and that a transverse current injected at one
 boundary  of the Hall bar, driven by an external electric field ${\cal{E}}$ gives rise to a
 local resistance that exponentially vanishes far from that boundary. Instead, the decay is power-law
 and there is a marked asymmetry between positive and negative values of the $B$ field, which is reduced at increasing  
 distance from the injection point.  Weak spin orbit coupling in graphene excludes that the  NLR is due to Spin Hall 
 Effect \cite{review_procolo,spin_hall_noi,Abanin2011a}.  The model of McEuen {\it et al.} \cite{McEuen1990} has been used by one of 
 us and collaborators to describe qualitatively and 
 semi-quantitatively the experimental results\cite{Nachawaty2018}. The model rests on the assumption that a non insulating bulk  plays a role in 
 transport  close to the CNP.  This is actually consistent   with the model  by Abanin  {\sl et al.}\cite{Abanin2011a}, which we  review 
 in Section III and accounts for some dissipative effects measured in graphene at the CNP by assuming  leakage of edge carriers
 into the bulk.  The $\nu=0$ Landau level would be splitted by Zeeman interaction and a sort of plateaux in the $xy$-conductance
 would survive at the CNP accompanied by a sizeable longitudinal  bulk resistance.  Indeed, 
 graphene on SiC,  in which QHE was measured with a metrological precision,  is  relatively disordered, with strong intervalley scattering.
 However, these models are unable to catch 
 the  asymmetry under a change in sign of the applied  the magnetic field.
 
 In this work  we have reconsidered  the possibility that dissipation in the bulk can be accompanied by local charge imbalance
 and  thermal gradients diffusing across the Hall bar, thus being responsible  for the NLR.  We can expect that even in the presence 
 of relatively strong magnetic field, close to the CNP, in the high-mobility bulk graphene, the Fermi surface vanishes, leading 
 to ineffective screening and  linear energy dispersion, which implies the formation of a strongly-interacting quasi-relativistic 
 electron- hole Dirac fluid \cite{Sheehy2007}. 
 
 In the relativistic approach, which is reviewed in Section II, particle current with energy transfer can be a neutral excitation 
 mode  independent of the charge transfer. The linear energy dispersion is also responsible for  long   recombination relaxation 
 times of the quasiparticles, so that local charge imbalance can persist during the diffusion. We propose that the 
 McEuen model accounts for the large part of the NLR, but long lived thermal effects occurring at low temperature in the bulk of 
 the Hall bar may add a relevant contribution which, in particular, could be the source of the asymmetry when flipping $B$.  In fact  the McEuen model requires 
 bulk conductivities $\sigma _1 ,\sigma_2$  for particles and holes respectively, which are not the same.
 
 On these side effects we focus on.  At low temperature ($T\sim 1\: ^oK$), with local non 
 equilibrium induced by a locally applied transverse electric field,  a few-Tesla  $B$    induces space separation of oppositely 
 charged carriers   leaking from opposite  edges, with little  recombination relaxation rate.  
 The charge imbalance  generated  in the bulk of the Hall bar by Joule heating close to the source is drifted far from the 
injection point by  a  
 Nernst thermal gradient  at open circuit.  In our framework, this is monitored   by the space dependence of 
 the imbalance chemical potential  which does not decay exponentially along the Hall bar  over distances comparable with the
 sample length. The bulk charge imbalance accumulated at opposite edges turns into a non local voltage that can be probed by 
 an external transverse current  which produces the NLR.  The contribution discussed here 
 appears in Fig.(\ref{vconb}) and is strongly asymmetric under changing the sign of $B$.
 
 Various remarks are in order. 
 
  As mentioned at the end of Section VI, we assume that the applied perturbation is localized close to the origin and only the 
  drift of charges induced by the thermal gradient is accounted for   in solving the diffusion equations derived in the hydrodynamical limit, 
  far away from the injection point. Self-consistency is invoked   to fix the non local voltage in the absence of sources.  
  This implies that, far from the source,  the set of differential  equations  is   homogeneous and the solution is up to an unknown 
  constant energy scale $ k_B \overline{T}_0$. We have tried to estimate this constant in conjunction with the Ettingshausen parameter, 
  but it can be of course  fitted from experiment.  Once the  polarization of the electric field ${\cal{E}}$   is fixed,  
  the response of the system to opposite directions of the applied $B$ field  is certainly asymmetric, as particles are exchanged 
  with holes when flipping the orientation of the magnetic field. The asymmetry is stronger close to 
  the injection point.  The  asymmetry with $B$ found experimentally can help in fixing the  ratio between 
  the contribution coming from Eq.(\ref{Ahallvolt6}) and the one coming from the McEuen model. 
  
  The charge imbalance arises from the spatial separation  of charges, as it can be seen from  the sketch of Fig.(\ref{scheme3}). In fact,  
  the magnetic field tends to keep carriers of opposite sign far away from one another. This reduces the chances  of recombination.  
  However some charge imbalance also originates from the difference between the bulk conductivities  $\sigma _1 ,\sigma_2$ of the 
  opposite charge carriers, which could be  the source of a reduction of the Coulomb interaction between charges.  This difference 
  parametrized by the variable $q$ of Eq.(\ref{qbet}) increases the asymmetry with the magnetic field orientation  as $q$ is odd with $B$ 
  and appears explicitely on the r.h.s. of Eq.(\ref{Ahallvolt6}). It is remarkable that, according to our model,  the closer the bulk 
  conductivities are, the longer is the  decay length $\lambda $ of the NLV, as can be seen from Eq.(\ref{length}).
 
  We assume that the Dirac model, in presence of two carriers of opposite charge  (at the CNP), is the  appropriate  low energy framework to discuss  
 the thermal, as well as  the charge transport. Hence we only consider the lowest bath temperatures attained 
  in the experiment which are $T \sim 1\div 1.7 \: ^o\!K $. In Eq.(\ref{mui})  an estimate is given of the charge imbalance chemical potential 
  $ \delta \mu _I $ in units $ k_B \: ^o\!K $  and is comparable with the bath temperature. Being estimated  close to the origin, where 
  the perturbation acts,  $ \delta \mu _I \sim 2 $   appears as the maximum attainable value. 
 Hence we are at the limit of validity for our approach,  because, in this range of values, the Dirac model  gives up by  turning into 
 the  more appropriate  Fermi  Liquid model. In  passing, we note that, at these temperatures, similar, but $p$ doped, graphene/SiC  samples have been used as barrier for Al/graphene/Al superconducting Josephson  junctions, showing a remarkable hysteretic loop of the critical current when  the magnetic field is cycled and an Aslamazov-Larkin (AL) paraconductivity  temperature dependence consistent with a Fermi Liquid picture\cite{Massarotti2016,Massarotti2017}.

   In the Dirac model the  relativistic stress energy tensor  provides  the energy flux $ T^0_\alpha $ from the  four-momentum conservation. 
   The full approach does not contain more general dissipative terms except in Eq.(\ref{Ahallvolt6}) for $v_{y} $, 
   in which a phenomenological dissipative term has been added. The recombination time $ \tau _Q \sim 1/\sqrt{B}$ has been measured
   and found rather long\cite{Rana2007,Winnerl2011}. This extra term has been added to account for the relatively fast decay of $v_{y} $ 
   with increasing $B$. If we neglect this term, the decay is determined by $  \lambda $ given by Eq.(\ref{length})
   which  gives a weaker dependence on $B$. 
   
  The numerical values for the electrical bulk conductivities are derived from Ref.[\onlinecite{Nachawaty2018}], 
  while the numerical values chosen for the thermal  transport 3-d parameters (Nernst  and  Peltier parameters, 
  thermal conductivity) have been found in the literature. In various occasions we have compared those values with the 
  ones derived from 3-d hydrodynamics of a Fermi Liquid\cite{Mueller2008a,Mueller2008b}.  This has given us the chance 
  of fixing the order of magnitude of  two quantities that are relevant in thermal transport, but are difficult to extract from experiment:
  the enthalpy  per particle $ e Q $ and the particle density in the effective 3-d system $ n_{3d} $.  Thermal 
  transport is an essentially 3-d phenomenon and the knowledge of the sheet  carrier  density  of graphene
  $ n_\square$, as determined by the QHE, is not enough.  The value we have chosen for  $ e Q $ is not far from 
  $k_BT$ plus the magnetization work $ -M\: B $ which  freezes the kinetic energy term. This reduces $\lambda $ 
  with increasing $B$. 
  
  Our results cannot be valid in the limit of small magnetic field ($B \lessapprox  1\: Tesla$).  Outside of the QHE  
  regime the model, in the absence of edge modes, is invalid from the outset. However the trend  when the magnetic 
  field is reduced is qualitatively well represented by our results of Fig.(\ref{vconb}).  Of course the transport
  parameters should be recalculated  in a Dirac frame and in presence of QHE quantization of the conductivity. 
  Nevertheless, hydrodynamics is still valid at very low energy (very low temperatures), lower that the Zeeman 
  splitting at $\nu =0$.  The starting guess is that the  bulk of the Hall bar is not totally insulating even at such a small  working temperature. 
  
  We conclude by noting that further studies of the  thermal transport in chemical vapour deposited graphene can help by making 
  the fractional quantum Hall effect an efficient  and controlled playground for  applications in alternative quantum information processors. 
 
\begin{acknowledgments}
	% put your acknowledgments here.
	We thank P. Brouwer, A. De Martino  and  R. Egger for fruitful discussions. This work was founded by project PLASMJAC Univ. 
	Napoli "Federico II".
\end{acknowledgments}

\begin{appendix}
    \begin{widetext}

%\section{Appendix}
%\setcounter{equation}{0}

\section{Diffusion equation for the average  temperature $ \overline T$}

      Here we derive  the diffusion equation for  the temperature $\overline{T} (x)$, induced by the imbalance charge $ \rho_I (x)$ defined in Eq.(\ref{imbo}),
      $ \rho_I =  n_\square \left (c_{+,A}-c_{+,B}-2\: s_+  \right )$.

 Using the notation introduced in Section III,  We assume the  mirror symmetry between the  side
 $A$ and $B$  of the bar, so that   $s_{+}  =   s_{+,A}= -s_{+,B} $ in the following. 
 
       Solving Eq.s(\ref{ureqt1}) with respect to the potentials $ "s" $,
       we get  ($Q \equiv\left (  \frac{ \epsilon +{\cal{P}} } {n\: e} \right )_{3d} $)

 \bea
   s_{-,A}-s_{-,B}  = - K_- \: \left ( -\partial _x\Delta  T \right ) -\frac{w}{4} \left ( 
   \frac{1}{\sigma_1}-  \frac{1}{\sigma_2} \right )\alpha _{xy}   \Delta T \:  \frac{{\cal{E}}\:  }{Q} 
 -\frac{w}{4}  \left (  \frac{1}{\sigma_1}+ \frac{1}{\sigma_2} \right )  g\:\frac{2\gamma}
{2\gamma+g }
\: (s_{-,A}+s_{-,B} )
   \label{eqp1}\\	
 s_{-,A} +s_{-,B}  =  - K_{-} ( -\nabla _x \overline{T} )-\frac{w}{4 }\: \left (  \frac{1}{\sigma_1}-  \frac{1}{\sigma_2} \right ) 
 \:\alpha _{xy}  \overline{T} \:  \frac{{\cal{E}}\:  }{Q} 
 - g\:\frac{2\gamma}
{2\gamma+g }
\:   \frac{w}{4 }\: \left (  \frac{1}{\sigma_1}+  \frac{1}{\sigma_2} \right )\:  \left ( s_{-,A} -s_{-,B}  \right ) 
 \label{truq} \\
2 s_+  =  - K_+ ( -\nabla _x \overline{T} )- \frac{w}{4} \left (  \frac{1}{\sigma_1}+ \frac{1}{\sigma_2} \right ) \: \alpha _{xy} 
\overline{T} \:  \frac{{\cal{E}}\:  }{Q} 
-\frac{w}{4}\left (  \frac{1}{\sigma_1}-  \frac{1}{\sigma_2} \right )
 g\:\frac{2\gamma}
{2\gamma+g }\:  \left ( s_{-,A} -s_{-,B} \right ) \; , 
\label{duqa}
 \enea 
where
 \beq
  K_\pm = \frac{w}{4}\left [\left (  \frac{1}{\sigma_1}\pm  \frac{1}{\sigma_2} \right )  
  \frac{\kappa _{xx} }{Q}+\left (  \frac{1}{\sigma_1}\mp  \frac{1}{\sigma_2} \right ) \: \alpha _{yx}\right ].
  \label{kka}
\eneq
Singling out  the $-2 \partial _x s_+$ term from Eq.(\ref{cona}) and  substituting it into the  derivative of  Eq.(\ref{duqa}),  one gets:
\bea
  K_+ ( -\nabla ^2_x \overline{T} ) +\frac{w}{4}\left (  \frac{1}{\sigma_1}+  \frac{1}{\sigma_2} \right )  \alpha _{xy} 
  \nabla _x\left ( \overline{T} \:  \frac{{\cal{E}}\:  }{Q}\right )   +\frac{w}{4}\left (  \frac{1}{\sigma_1}-  \frac{1}{\sigma_2} \right )
 g\:\frac{2\gamma}
{2\gamma+g }
\: \partial _x  \left ( s_{-,A} -s_{-,B} \right ) \nonumber\\
 =   \frac{ \sigma _1 -\sigma _2 }{ \sigma _1 +\sigma _2 } \:  \partial_x\left (s_{-,A}+s_{-,B}\right ) 
 -\frac{  2  \: v_F \:w}{(\sigma _1 +\sigma _2 )} \: \frac{1}{n_\square{\cal{V}}} \rho_I. \nonumber
 \enea
 In the comparison with the derivative of  Eq.(\ref{truq}): 
 \bea 
  \partial _x  \left ( s_{-,A} +s_{-,B} \right ) = - K_- ( -\nabla ^2_x \overline{T} ) -\frac{w}{4 }
  \left (  \frac{1}{\sigma_1}-  \frac{1}{\sigma_2} \right )\alpha _{xy}  \nabla_x\left ( \overline{T}  \frac{{\cal{E}}  }{Q}\right )  -   g\:\frac{2\gamma}
{2\gamma+g }
 \frac{w}{4 }\: \left (  \frac{1}{\sigma_1}+  \frac{1}{\sigma_2} \right )   \partial _x \left ( s_{-,A}-s_{-,B} \right ), 
 \label{ff47}
\enea
the contribution from $\partial _x( s_{-,A}-s_{-,B})$ cancels out, giving eventually the final form of the equation we are looking after:
  \bea
   \kappa _{xx}\:( -\nabla ^2_x \overline{T} ) +   \alpha _{xy}  \nabla _x \left ( \overline{T} \:  {\cal{E}} \right )= - 4 \: 
   \left (  \frac{\epsilon + {\cal{P}} }{n\:  e^2}\right )_{3d}\: \frac{1}{n_\square {\cal{V}}} \:  v _F \:
   \frac{   \sigma _1\sigma _2 }{(\sigma _1^2 +\sigma _2 ^2)} \: \rho_I,
  \label{Aeqpx}
 \enea
 \end{widetext}
 which can be cast in the form ($ \theta  \equiv  {\alpha _{xy}  \:  {\cal{E}} \: L }/{  \kappa_{xx} } $):
 \bea
\partial^2_{\tilde x}  \tau  - \theta \: \partial_{\tilde x}  \tau =   \tilde\beta \delta\mu_I,  
 \label{Apert} 
     \enea
 %   \begin{widetext}
    \bea
 \tilde \beta  = \left (  \frac{\epsilon + {\cal{P}} }{n\:  e^2}\right )_{3d}\:  \frac{  v _F}{n_\square {\cal{V}}} \: 
 \frac{k_B }{\kappa _{xx}} \:\frac{  4 \: \sigma _1\sigma _2 }{(\sigma _1^2 +\sigma _2 ^2)}
   \times  \left [ 1- \left (  \frac{\sigma _B }{ \sigma _{3d} }\right )^2 \right ] ^{-1}  \: ,  \nonumber\\
     \label{betn}
   \enea
 %   \end{widetext}
where $\delta\mu_I =  \frac{e \rho_I L^2  }{k_B \: ^oK} $ and  $\tau =  \overline{T} /\: ^oK $. 
 %   \appendix

      \section{Equation for the imbalance  energy  $\delta\mu_I $}   
       
    We now  derive the equation that relates the imbalance energy   $\delta\mu_I $  to the average temperature $\overline T $. 
    % $\delta\mu_I $  is defined as $\delta\mu_I =  \frac{e \rho_I L^2  }{k_B \: ^oK} $, where  $\rho_I$ is the imbalance  charge density, 
     From Eq.s(\ref{typea}), we get: 
      %and by defining   $  2\:  s _{+} = s_{+,A}-s _{+,B} $,   to simplify the approach, we get
              \begin{widetext}
 \bea
  \partial_x  (c_{+,A}- c_{+,B}) & = - ( 2\: \gamma '+g') ( c_{-,A}+ c_{-,B})+ g '( s_{-,A}  -  s_{-,B} ) \label{ff1}\\
   \partial _{x} (c_{-,A}-c_{-,B}) & =  -g' \:  ( c_{+,A}+c_{+,B}) \label{ff2}\\
  \partial_x  (c_{+,A}+ c_{+,B}) & = -( 2\: \gamma' +g') ( c_{-,A}- c_{-,B})+ g' (  s_{-,A}  + s_{-,B} ) \label{ff3}\\
 \partial _x \left ( c_{-,A} + c_{-,B} \right ) & = - g'\: \left [ c_{+,A} - c_{+,B} - 2 \: s_{+}\right ] \label{ff4}\\
\nonumber \\
 \partial^2 _{x} (c_{-,A}-c_{-,B}) & =  g'\: (2\gamma '+g') ( c_{-,A}-c_{-,B}) - {g'} ^2\:  (s_{-,A}+s_{-,B})\nonumber \\
  \partial^2 _{x} (c_{+,A}-c_{+,B}) & =  g'\: (2\gamma' +g') \left [  c_{+,A}-c_{+,B} -  2 s_{+} \right ]+ g' \:  \partial _x  (s_{-,A}-s_{-,B}) 
  \label{Atypeass}
        \enea

Using  Eq.(\ref{Atypeass}) and Eq.(\ref{cona}), we also obtain:
    \bea
 \partial _x \left (c_{+,A}-c_{+,B}-2\: s_+  \right )=  - (2\gamma +g ) \: t_0\:  (c_{-,A}+c_{-,B}  ) + g\: t_0\:  \left (s_{-,A}-s_{-,B}  \right ) \nonumber\\
 -\frac{  2\: w  \:  v_F }{( \sigma _1 +\sigma _2 ) {\cal{V}}} \:  \left (c_{+,A}-c_{+,B}-2\: s_+  \right )
 +\frac{ \sigma _1 -\sigma _2 }{ \sigma _1 +\sigma _2 } \:  \partial_x\left (s_{-,A}+s_{-,B}\right )
   \label{con3}
   \enea
The parameters $g$ and $\gamma $ have dimensions $[ 1/( L t_0 ) ]$ where $t_0$ is the relaxation  time scale  
that we have introduced in the definition of the dimensionless leakage  parameter from the edge states to the 
bulk, $\tilde g = g( 2\:\gamma +g )L^2 t_0^2$  to properly match the dimensions  
( see below for details).
         By differentiating Eq.(\ref{con3}) and by plugging into it  Eq.s(\ref{eqp1}), 
    keeping terms up to ${\cal{O}} ( g^2) $,  one recovers the desired equation, connecting the imbalance potential with 
    the thermal excess temperatures, given by:
         
         \bea
     \partial ^2_x  \left (c_{+,A}-c_{+,B}-2\: s_+  \right ) +\frac{  2\: w \: \: v_F }{( \sigma _1 +\sigma _2 ){\cal{V}}} \:   \partial _x
\left (c_{+,A}-c_{+,B}-2\: s_+  \right )  - g \:  (2\gamma +g ) \: \tau _Q^2\left (c_{+,A}-c_{+,B}-2\: s_+  \right ) \nonumber\\
+\frac{ \sigma _1 -\sigma _2 }{ \sigma _1 +\sigma _2 } \: \partial_x^2
  \left \{  K_{-} ( -\partial_x \overline{T} )+\frac{w}{4 }\: \left (  \frac{1}{\sigma_1}-  \frac{1}{\sigma_2} \right ) \:\frac{ \alpha _{xy} 
  \: {\cal{E}} }{Q} \:  \overline{T} \right \} \nonumber\\
  =  -g \: \tau_Q  \left [ 1 - \frac{2\gamma\: \tau_Q^{-1}}{2\gamma+g }
\: \frac{w}{4 }\: \left (  \frac{1}{\sigma_1}+  \frac{1}{\sigma_2} \right )\: \frac{ \sigma _1 -\sigma _2 }{ \sigma _1 +\sigma _2 } \right ] \: 
\partial_x 
  \left \{ K_- \left ( -\partial _x\Delta  T \right ) +   \frac{w}{4} \left (  \frac{1}{\sigma_1}-  \frac{1}{\sigma_2} \right ) \:\frac{\alpha _{xy} 
  \: {\cal{E}} }{Q} \: \Delta T \right \},
     \label{con4}
   \enea
  Note that the expression in curly brackets is the same, both for $\overline T $ and for $\Delta T $.  One can assume that one of the two is
  constant.  
   We neglect any $\hat{y}-$dependence of $\rho_I$ except for the mirror symmetry,  so to make it independent of  $\Delta T = T_A -T_B$.  
   Hence the right hand side is put equal to zero:
   \bea
     K_- \left ( -\partial _x\Delta  T \right ) +   \frac{w}{4} \left (  \frac{1}{\sigma_1}-  \frac{1}{\sigma_2} \right ) \:\frac{\alpha _{xy}  
     \: {\cal{E}} }{Q} \:   \Delta T   = cnst, 
     \label{cono}
\enea 
where the constant can be put to zero. Our  choice is consistent  with the initial choice   $s_{+,A} = - s_{+,B}= s_+ $.
                   Finally,  multiplying both sides by  $ e\: n_\square L^2 / k_B  \: ^oK $ and resorting  to 
                   the dimensionless space coordinate $\tx = x/L$,  we get:   
           \bea
      \partial ^2_{\tilde{x}} \delta\mu_I  +\frac{  2\: w \: L \: v_F }{( \sigma _1 +\sigma _2 ){\cal{V}}} \:   \partial _{\tilde{x}}
\delta\mu_I    - \tilde{g} \: \delta\mu_I  
 =- \frac{e}{k_B} \frac{n_\square L^2}{Q}  \: \frac{ \sigma _1 -\sigma _2 }{ \sigma _1 +\sigma _2 } \: \frac{w}{4\: L} \:  \partial^2_{\tilde x} 
 \left [\left \{\left (  \frac{1}{\sigma_1}-  \frac{1}{\sigma_2} \right ) \:  \kappa _{xx}   
\right .\right . \nonumber \\
\left .\left . +\left (  \frac{1}{\sigma_1}+  \frac{1}{\sigma_2} \right ) \: \alpha _{yx}\: Q  \right \} ( -\partial_{\tilde x} \tau)
 +\left (  \frac{1}{\sigma_1}-  \frac{1}{\sigma_2} \right )  \:  \kappa _{xx} \: \left ( \frac{\alpha _{xy} }{ \kappa _{xx}}\: {\cal{E}} \: L   
 \right ) \: \tau\right ]. \label{Apermu}
  \enea
   \end{widetext}  
      
\section{ Fermi liquid transport parameters}

 The scattering time is the $ e-e$ inelastic one, because this is the time scale which qualifies the  imbalance relaxation rate: 
 $\tau_{ee} ^{-1}=  \alpha_f ^2 \: k_B T / \hbar $. The elastic scattering is unable to relax the charge  imbalance.  At $B = 2 \:Tesla$,
 one gets: 
 
\bea
 \omega _c \tau = \frac{e B}{c }\frac{v_F}{\hbar k_F } \tau_{ee}  = 0.4626 \times 10^{6} \: ,
\nonumber
 \enea
with  $k_F = ( \pi \: n_{\square} )^{1/2} $ and  $   n_{\square}  = 10^{10}/ cm^2 $, $ v_F = c/ 300$, $B= 2 \: Tesla$.

 In terms of the enthalpy, the thermal response functions in the limit $ \omega _c \tau >>1 $ are given by Eq.s(\ref{set},\ref{seit}). In  
 the Fermi liquid regime, in which  the thermal energy is lower than the chemical potential, the  expressions for the transport coefficients 
 $\alpha _{xy}^{FL}$ and $\kappa _{xx}^{FL}$ are:
\bea
\alpha _{xy}^{FL} =   \left( \frac{ \epsilon + {\cal{P}} }{ n \: e} \right )_{3d}\: \frac{\sigma _B}{T}.\label{alpqui} \nonumber\\
\kappa _{xx}^{FL} \approx   \left( \frac{ \epsilon + {\cal{P}} }{  n\:e  } \right )_ {3d}^2\frac{\sigma _{3d}}{ T }  \left [ 1 - 
\left ( \frac{\sigma _{B}}{ \sigma _{3d} }\right )^2 \right ].
\label{kap}
\enea
 These are quantities defined in 3-d, where  their 3 dimensions  are
 respectively given by $[\alpha ]\sim  e/ (L t \: T) $   and by  $[\kappa ]\sim  e^2/ ( L^2\: t \: T) $. 
 On the other hand, in the range  $T= 1\div 10 \:  ^o\! K $  the value reported for graphene is \cite{Checkelsky2009} $\alpha _{xy}  
 \approx 75 \: n  Amp /  ^oK $, which is a 2-d quantity. As we consider thermal effects in the graphene grown onto SiC as a "bulk " 
 phenomenon, we extend the definition of  $\alpha _{xy} $ to $3d$ by dividing it by $ \sim0.3 \: nm$, so that we eventually obtain:
 \beq
  \alpha _{xy}  \approx 2.24 \times  10^{11}  V/ ( ^oK \: sec) .
  \label{alpquo}
  \eneq
 Ref.[\onlinecite{Mueller2008b}] provides a numerical  estimate  $ \left(\epsilon + {\cal{P}}\right )_ {\square}  \sim 0.12 \times 10^3 \:
 (T/\: ^oK )^3\: eV/ cm^2$, also given in Ref.[\onlinecite{Crossno2016}], when the chemical potential exceeds  thermal energy, even  at 
 temperature  $40 \:^oK$. The question is how to relate this quantity to the corresponding 3-d quantity$ \left(\epsilon + {\cal{P}}\right )_ {3d }$, 
that is required for the enthalpy  per unit (3-d) volume.
 
% There is an estimate of this quantity\cite{mueller2}  but we are unable to tell what is the substrate considered in that work, nor what is  the proportionality with $  Q\equiv \left ( \frac{\epsilon + {\cal{P}} }{ n\: e}\right  )_{3d}$, which should be considered for our special case of impurities in SiC. To solve this ambiguity, we put both quantities equal. In turn, this position fixes the relation between $n_\square$ which is well known from Hall measurements, $n_\square  \sim 10 ^{10} /cm^2 $,  and $n_{3d} $ which is unknown and can be expressed in terms of the 'unity'  volume  $   {\cal{V}} = [\ell w d]$ introduced  in  Eq.(\ref{equa}).

%\sim \left(\epsilon + {\cal{P}}\right )_ {\square}  / 0.3 \: nm $ gives, from Eq.(\ref{alpqui}), $ \alpha _{xy} \sim 3  \times 10^6  V/ \: ^oK \: sec $, at   $T= 1 ^oK $, which is about five  order of magnitudes less than  the value reported in Eq.(\ref{alpquo}). 
%Of course, we are almost  in the relativistic regime where the temperature exceeds  the chemical potential, though  and the two  order of magnitude do not have to match.

 We make the following choice.  We  define  a total energy per particle  by multiplying $ \left(\epsilon + {\mathcal{P}}\right )_ {3d }$ 
 by the volume ${\mathcal{V}} = \ell w \times 1 \AA  \approx 10^{-11} cm^3$ and dividing it by  the  effective number of 3-d carriers  
 ${\cal{N}}_{3d} \sim 10^3$ and we pose:
\bea
Q=  \frac{\left ( \epsilon + {\cal{P}} \right )_{3d}{\cal{V}} }{ {\cal{N}}_{3d}\: e}   \approx   0.12 \times 10^{-3} \: V 
\label{ququ2}
\enea
 (here and in the following  $V\equiv Volt$).   We will show that this value for ${\cal{N}}_{3d}$ is also suggested by our self consistent procedure. 
 $   {\cal{N}}_{3d} $ can be interpreted as  the effective  number for carriers contributing to the thermal processes in the 3-d strip.
 This  value for ${\cal{N}}_{3d}$ allows us to match reasonably well the values  from a Fermi Liquid approach of Eq.(\ref{kap}) and those 
 experimentally found for $\alpha _{xy}$, $\kappa _{xx}$ and $ e_N$.  The value reported by experimentalists for  $\kappa _{xx} $,adopted by us,  
 is $\kappa _{xx}  \approx 1 \div 10 \: Watt / ( ^oK \times meter)$ . 
 
 The ratio  $   \kappa _{xx}^{FL} / \alpha _{xy}^{FL} $ is given by:
\bea
  \frac{    \kappa _{xx}^{FL} }{  \alpha _{xy}^{FL}}  =   \left( \frac{ \epsilon + {\cal{P}} }{ n \: e} \right )_{3d} \frac{\sigma _{3d}}{\sigma _B}  
  \left [ 1 - \left ( \frac{\sigma _{B}}{ \sigma _{3d} }\right )^2 \right ], 
  \label{ratiov}
  \enea
    As $ \sigma _{3d}  L ^{d-2}  = R^{-1} $, an aspect ratio is required, so that  $ \sigma _{3d}  ^{-1} = R\ell d / w  
    \approx 0.5 \times 10^{-15} \: sec$, with     $d =1 \AA $ and $R=10\:k \Omega $. Here $R$ is the longitudinal resistance and $ \ell w$ the area. 
    % In fact,   $h/ (e^2 \: 10\: k \Omega) \sim \frac{ 4}{\pi} \frac{ 20. 273 k\Omega }{10\:k \Omega } \sim  2.6$. 

 Here  $\sigma _B/\sigma _{3d} = e^2R /h \cdot [ hc /( eB\ell w)]\cdot \ell /w $(note that, as it happens  in the hydrodynamic 
 approach, the limits $ B \to 0 $ and  $\omega \to 0 $  do not commute).   We find
  \bea
       \sigma _B =  \frac{ {\cal{N}}_{3d}}{ {\cal{V}}} \frac{e\:c}{B} \approx  2.4 \times \frac{10^{15}}{cm^3}. \frac{ ec}{B}.\nonumber\\
        = 0.84 \times 10^{12} sec^{-1} 
       \enea
  With these conductances, the ratio of Eq.(\ref{ratiov}) is estimated  as:
  \bea
   \frac{    \kappa _{xx}^{FL} }{  \alpha _{xy}^{FL}}  \approx 4.\times 10^{-2}  \: V
   \label{ratio}
\enea 
The Nernst coefficient  in the limit of large fields and relatively clean samples\cite{Checkelsky2009}  is $ e_N^{3d}  =   (\sigma ^{-1}   
\alpha ) _{xy} \approx  1 \:  m V / ^oK $. 
 
% Implicitely, for $n_\square = 10^{10} / cm^2 $, this choice provides $ n_{3d} \approx  10^{11}/cm^3$.

%Q=  \left( \frac{ \epsilon + {\cal{P}} }{ n \: e} \right )_{3d} \sim \left ( \frac{ \epsilon + {\cal{P}}  }{ n\:  e}\right )_{\square}  \frac{n_\square}{n_{3d}} \nonumber\\ \approx  0.12  \times \left ( \frac{T}{1 \: ^0K} \right )^3 \times  10^{-3} V, 

%which is quite high, but is not meaningful in our quasi $2d$ system and we will never rely on it anyhow. Our choice assumes good metallic thermal transport in the system.

 %Defining $n_{3d} =  {\cal{N}}_{3d}/  {\cal{V}}$.
 
% We will find  from consistency  that 
     
  %   \frac{ \rho _I}{e} = q \frac{ {\cal{N}}_{3d}}{ {\cal{V}}} \:\:\: with\:\:\:   \frac{ {\cal{N}}_{3d}}{ {\cal{V}}}  = 2.4 \times \frac{10^{15}}{cm^3}.
   %   \enea
     % With this density, at $B = 2 \: Tesla$,

      \section{Coupled differential equation and relaxational scales }
      
      Let us rearrange the system of differential equations involving $ \bar T$ and $ \delta \mu_I $, Eq.(\ref{Apert})  
      and Eq.(\ref{Apermu}), in a more practical form so to be able to  discuss the  relaxation length scale of the two unknown 
      functions in the $\hat{x} -$direction. 
      
      We write the prefactor of  $ ( -\partial_{\tilde x} \tau)$ as $  k \equiv   u\: k_u$  where  $k_u$ is given by:
      \bea
    k_u = 1 +\left (  \frac{\sigma_1+  \sigma_2}{\sigma_2-  \sigma_1} \right ) \: \frac{\alpha _{yx}}{\kappa _{xx}} \: Q ,
         \label{ku}
    \enea
    and  $u$ is given by:
     \bea 
u=  \frac{e}{k_B} \frac{ n_\square L^2}{Q} \frac{ \sigma _1 -\sigma _2 }{ \sigma _1 +\sigma _2 } \: \frac{w}{4\: L} \: 
\left (  \frac{1}{\sigma_1}-  \frac{1}{\sigma_2} \right ) \:   \kappa _{xx} 
\label{uu}
  \enea

    Eq.s(\ref{tau1}) follow: 
 \bea
 \partial  ^2_{\tilde x} \: \tau  -  \partial _{\tilde x}\: \left ( \theta   \: \tau \right )  =   \tilde{\beta}  \:\delta  
 \mu_I, \:\:\: (a) \:\:\:\: \nonumber  \\
    \partial  ^2_{\tilde x} \:\delta \mu_I  + a\:  \partial  _{\tilde x} \:\delta \mu_I 
       - \tilde{g} \: \:\delta \mu_I  =\partial  ^2_{\tilde x} \: 
  \left \{ k \: \partial _{\tilde x} \: \tau - \theta \: u  \:\tau  \right \} .  \:\:\: (b)  \nonumber \\
  \label{Atau1}
  \enea
   A numerical solution is obtained by deriving an equation for $\delta \mu_I $ only and by inserting the solution into Eq(\ref{Atau1},(a)).  
   Multiplying  Eq.(\ref{Atau1},(a)) by $k$  and substituting it in Eq.(\ref{Atau1},(b)) ,  we get:
\bea
(k- u ) \:   \partial ^2 _{\tilde x}  \left ( \theta \: \tau \right )  =   \partial ^2 _{\tilde x} \: \delta \mu_I + ( a- k \tilde\beta ) \:  
\partial  _{\tilde x} \: \delta \mu_I -\tilde g \: \delta \mu_I \nonumber\\
\label{nonlin0}
\enea
From now on we shift  $ a \to a-  u \tilde\beta $ by summing and subtracting $u \tilde\beta $ to $a$, and ridefine it as $a$  thus obtaining:
\bea
(k- u ) \:   \partial ^2 _{\tilde x}  \left ( \theta \: \tau \right )  =   \partial ^2 _{\tilde x} \:\delta \mu_I + [ a- (k -u)\tilde\beta ] \:  
\partial  _{\tilde x} \: \delta \mu_I -\tilde g \: \delta \mu_I .\nonumber\\
\label{nonlin}
\enea
  We cast the system in length units  $a^{-1} L$: ($\tilde{\tx} =  x/(a^{-1} L)$  and introduce the new variables 
  $ \mu ' \equiv  \tilde{\beta}  a^{-1} \delta\mu_I $, $\tau  ' \equiv a\: \tau $ to get:  
  \bea
 \partial  ^2_{\tilde{\tx} } \: \tau ' -  \partial _{\tilde{\tx} }\: \left ( \theta a^{-1}  \: \tau' \right )  =  \mu ',  \:\:\:\:\: (a) \nonumber \\
    \partial  ^2_{\tilde{\tx} } \:\mu '  +  \partial  _{\tilde{\tx} } \:\mu  '
       - \tilde{g} a^{-2}\: \:\mu '  \:\:\:\:\:\: \hspace{4cm}   \label{verosys}\\
    =\partial  ^2_{\tilde{\tx} } \: 
  \left \{  ( \tilde{\beta}  k a^{-1}) \: \partial _{\tilde{\tx} } \: \tau '- \theta a^{-1}\: (  \tilde{\beta}  u a^{-1})  \:\tau ' \right \} . \: (b) \nonumber
  \enea
where 
\bea
 a \equiv  2\: \frac{ w }{( \sigma _1 +\sigma _2 ){\cal{V}}} \:  v_F ,\:\:\:\: \tilde{g} \equiv g ( 2{\gamma} +g ) \: L^2 t_0^2. 
  \label{kco}
 \enea
 The Fermi velocity $v_F= c/300 = 10^{8} \: cm/sec$.
   We take the parameter that describes the leakage from edge to bulk 
 $ \tilde g\sim 10^{-6}$, so that $\tilde g a^{-2} \sim 10^0$ . 

  For $\tilde{\beta} \sim 10^{-9}$ and $a\sim 10^{-3}$  Eq.(\ref{ku})  and Eq.(\ref{uu})  give ( see in the next,  Subsection D)
\bea
  k'\equiv  \tilde\beta k\: a^{-1} 
     =  q^2 + 3\times 10^{-3} \: q   \nonumber\\
 u'\equiv     \tilde\beta u\: a^{-1} =    q^2, \:\:\:\:\:  q = \frac{   \sigma _1-\sigma _2 }{[\sigma _1^2 +\sigma _2 ^2]^{1/2} }.\nonumber\\
     \label{bet}
\enea
 The order of magnitude of $\theta $ is fixed by the applied voltage (expressed in $Volts \:\equiv V$), as derived from the experiment: 
 \bea
  {\cal{E}} w \:  \sim 10^{-4}  V \to  R \sim  \frac{10^{-4}  V}{10 \:n\:  Amp} \sim 10  \: k \Omega
  \nonumber\\
  \to
 \theta ' \equiv    \theta\: a^{-1} = \frac{\alpha _{xy}}{\kappa _{xx} } \: {\cal{E}} w \:  \frac{  a^{-1} L}{w} \approx 3 \times 10^{-2}.
 \label{tep}
  \enea  
  The final equation is:
  \bea
   \partial ^2_{\tilde \tx} \mu ' +\left [ 1-( k' -u') -\theta ' \right ]   \partial _{\tilde \tx} \mu ' \nonumber\\
     +  \left ( g' + \theta '\right ) \mu ' - g' \theta ' \int \mu '  d  {\tilde \tx}  =0
   \enea

  \section{Properties of the imbalance charge energy solution}
  
 Having discussed the order of magnitudes and the general features of the solutions of the system   
 Eq.(\ref{Atau1}), we now  turn to a numerical solution  to highlight the dependence  
 of $\delta \mu _I(x)$ and $\tau (x)$ on the magnetic field $B$, both far from $x=0$ and close to the  origin. 
 
  %and assuming $b^{-1} \partial _\tx \tau \sim \tau $  as before, we find a relation between the   orders of magnitude of the kind: $ u \:a^{-2} \left [ b^3 \: k_u - b^2 \: \theta \right ] \tau  \sim \mu_I$ ( note that $u \: a^{-2} \sim 1$).
 
  We have shown in Appendix D that the system of Eq.(\ref{Atau1}) allows for a rewriting with parameters  and unknown functions of  $ {\cal{O}}(1) $. 
  The system of Eq.(\ref{verosys}) has been used in the numerical results which are discussed in Section V of the text. 
     %$\mu_I \sim $.   and, for  $ e {\cal{E}} w  \sim 10^{-4} eV$, is    $\theta \sim 6. \times 10^3$. 
     \bea
 \partial  ^2_{\tilde{\tx} } \: \tau ' -  \partial _{\tilde{\tx} }\: \left ( \theta ' \: \tau' \right )  =  \mu ',  \:\:\:\:\: (a) \nonumber \\
    \partial  ^2_{\tilde{\tx} } \:\mu '  +  \partial  _{\tilde{\tx} } \:\mu  '
       - \tilde{g} ' \: \:\mu '  \:\:\:\:\:\: \hspace{4cm}   \label{Atau2}\\
    =\partial  ^2_{\tilde{\tx} } \: 
  \left \{   k' \: \partial _{\tilde{\tx} } \: \tau '- \theta ' u'  \:\tau ' \right \} . \: (b) \nonumber
  \enea
   with $\tilde{g}' =  \tilde{g} a^{-2}$ and primed parameters given in Eq.(\ref{bet}, \ref{tep}). As discussed in the text, 
  the solution of the homogeneous equation Eq.(\ref{Atau2}, (b)) has to be added to the solution of the system of coupled equations. 
   
    Here we keep  the form of the system given in Eq.(\ref{Atau2}), but we drop all the primes for simplicity and we use $ \partial$
    instead of $\partial_{\tilde{\tx}}$.
       %$\mu_I \sim $.   and, for  $ e {\cal{E}} w  \sim 10^{-4} eV$, is    $\theta \sim 6. \times 10^3$. 
    
 %For $ \tilde g =0 $ a solution of  Eq.(\ref{Atau1},(a))  with $\mu_I = \: cnst $ is possible.  Then, the excess temperature takes the form given in Eq.(\ref{toy}).    For $ \tilde g \neq 0 $  and $\theta =0$, a solution  for the imbalance potential is of the form given by Eq.(\ref{teta0}) with an exponential decay   of the kind $ e^{ \tilde{g} \tilde{x} / F } $ with $ F\sim a$. 

                             For $ g, \theta  \neq 0$, a numerical solution is obtained by deriving an equation for $\mu $ 
                             only and, by inserting the solution into Eq(\ref{Atau1},(a)).   Multiplying  Eq.(\ref{Atau1},(a)) 
                             by $k$  and substituting it in Eq.(\ref{Atau1},(b)) ,  we get:
\bea
(k- u ) \:   \partial ^2  \left ( \theta \: \tau \right )  =   \partial ^2  \:\mu + [ 1- (k-u) ] \:   \partial  \: \mu - g \: \mu \nonumber\\
\label{nolin}
\enea
  Let us consider $  \partial \: \left ( \theta   \: \tau \right )  = \theta  \:  \partial \: \tau  + \tau \:  \partial \: \theta $. 
The second term on the r.h.s.   is reminiscent of  $   \partial _{\tilde x}\: {\cal{E}}_y -  \partial _{\tilde y}\: {\cal{E}}_x $ where 
$ {\cal{E}}_ x= 0 $ because, in the bulk, the circuit is open  ( no current along $x$ ) and assumed to be equipotential. 
This $z-$ component of $ \vec{\nabla}\times \vec{{\cal{E}}} $  vanishes, because there is no current in the $z$ direction. 
We conclude that only the first term survives with a $\theta $ which is practically constant in $x$, in an interval of size $\theta ^{-1}  L $ 
close to the origin, or zero for any $x$ far from the origin.  
 Following Eq.(\ref{Atau2}, \ref{nolin}), we define   
\bea
 \:\:\:\:  \:\:\:\:  \:\:\:\:  \nu (\tilde{x} ) =  \nu_0 +  \int ^{\tilde{x} }_0  \!\!\! \mu  (\tilde{x}' ) \: d \tilde{x}'
\label{eqq3}
\enea
and we get:
\bea
\left \{ \begin{array}{l}  
 \partial  ^2 \: \tau-  \theta   \:   \partial   \: \tau =  \mu ,  \\
  (k- u ) \:   \theta \:  \partial \: \tau   =   \partial  \:\mu + [1-(k-u)] \mu - g \: \nu \nonumber\\
       \partial   \:\nu = \mu 
        \end{array}  \right . 
\label{eqq5}
\enea
or, substituting the second into the first: 
\bea
\left \{ \begin{array}{l}  
(k- u ) \:  \partial ^2 \: \tau  =    \partial  \:\mu +  \mu - g \: \nu , \nonumber\\
  (k- u ) \:     \theta \:  \partial  \: \tau  =   \partial \:\mu + [1- (k-u)]  \:   \mu - g \: \nu \nonumber\\
       \partial   \:\nu = \mu  \end{array}  \right . 
\label{eqq6}
\enea

We now multiply the first equation by $\theta $ and differentiate the second one.  Equating  the two of them to each other, 
we obtain: 
\bea
\partial ^2 _{\tilde x} \: \mu  + (  a - k \:   \tilde{\beta} ) \:  \partial  _{\tilde x}\mu   - \tilde{g} \:  \mu  =   \theta  \:    
\partial  _{\tilde x} \: \mu  +  \theta \: \left ( a-  u \tilde\beta \right ) \: \mu   -\tilde{g}\:  \theta  \:  \nu . \nonumber
\enea  
        When $ \mu $ has been found, one can plug it into the first one  of Eq.s(\ref{eqq5}) to find $  \partial  _{\tilde x} \: \tau $. 
The equation  can be written as a first order differential system: 
\bea
\left \{ \begin{array}{ll}      \partial  _{\tilde x} \:\mu &= \xi \\
\partial  _{\tilde x} \:\nu &= \mu  \\   
 \partial  _{\tilde x} \: \xi  &= - [ 1   -(k - u ) - \theta ] \: \xi +[ {g}  +  \theta   ] \:  \mu  -{g} \: \theta \:  \nu  ,
       \end{array} \right .  \nonumber
         \enea
   or,  in the matrix form,
          \bea
\left ( \begin{array}{c}    \partial \:\mu \\ \partial  \:\nu \\  \partial  \: \xi \end{array} \right )  =  \left ( \begin{array}{ccc} 
0 & 0  & 1 \\ 1 & 0 & 0 \\ G &   -{g}\theta \:  & -F \end{array} \right ) \left ( \begin{array}{c}  \mu \\ \nu \\ \xi \end{array}\right ) 
\label{matta}
         \enea  
          with  $ F =    1  -(k - u ) - \theta  $ and $ G =   {g} +   \theta $.   
          As discussed above,  $ a \to a a^{-1} \sim 1 $, $ k\tilde{\beta} \to k \tilde{\beta} a^{-2} $, $ \tilde{g} \to  \tilde{g} a^{-2} $, 
          $ u \tilde{\beta} \to  u\tilde{\beta} a^{-2} $ and $\theta$ are all $ {\cal{O}}(1) $.
          % Here $\tilde{\tx} \to \tx$ to make notation simpler. 
           The characteristic equation  giving the space dependence  of the functions here $ e^{\tilde{\lambda} \tilde{x}}$ is
          \bea
           \tilde{\lambda} ^3 + F \:  \tilde{\lambda} ^2  - G \:  \tilde{\lambda} + \tilde g \: \theta =  0 
           \label{cub}
           \enea
          % For $\theta = 0 $, the space translational invariant solution $ \tilde{\lambda} =z_3=0$ is available. This is unphysical and should be dropped, as it arises from the approximation of taking $\theta $ independent of $x$.
         %  Given $ z_3 = 0$, it is easily found that $ z _{1,2} = - \frac{F}{6} \pm  i\: \sqrt{- \tilde g } $,  with $\tilde g  <0 $,  so that in the far zone, $\theta =0$, is $\gamma  \approx \sqrt{- \tilde g } $  ( see  Eq.(\ref{gamma}) ). Purely oscillating solutions (i.e.  purely imaginary) would require  $F \to 0$, which is excluded. 

   For $\theta \neq 0 $,  solutions converging at infinity require $\tilde g ,\theta <0$.
                      
                        The vanishing of  the imbalance parameter   $q$ implies the vanishing of $(k - u )$. Here we show  that, 
                        for $\theta \neq 0 $, the  vanishing of $q$, provides two real or complex conjugate solutions for 
                        $  \tilde{\lambda} $  which are independent of $\theta $ and a real solution  $ z_3 = - \theta\:  sign[\tilde{g}] $. 
                       
                     If $z_1,z_2,z_3$ are the roots of the cubic equation the following relations hold:
     \bea
      z_1z_2+ z_1z_3+z_2 z_3 = -G,\nonumber\\
      z_1+z_2+z_3 = -F, \:   z_1z_2z_3 = -\tilde{g} \theta .
        \label{trerv}
      \enea
Given $ z_{1,2} =  -a/2  \pm \sqrt{(1 - 4 |{g}| }/2 $, it is easily found that  $ z_3 = - \theta\:  sign[{g}] $ satisfy Eq.s(\ref{trerv}). 
When $q\neq 0$,   one solution will be real and the other two will be complex or all of them will be real. 
%In search for purely oscillating solutions with no decay rate at $\theta \neq 0 $, the  first equality in Eq.( \ref{trere}) implies  $  z_2, z_3 =\pm i\: \sqrt{ -G}$ with  $- G> 0$. Rewriting the cubic equation in factorized form and comparing equal powers with  Eq.(\ref{cub}) :
 %   \bea  (\tilde{\lambda} - z_1) ( \tilde{\lambda }^2 -G) =0,       \freccia  z_1 = - F,\:\: \:   -F\: G =  \tilde g \: \theta. \nonumber \enea
% Purely oscillating solutions with no decay in space  at $\theta \neq 0 $ would require $F\: G =  -\tilde g \: \theta $, which may only occur accidentally for some given  value of the magnetic field,   even when  dropping terms  ${\cal{O}} [ {\cal{E}} ^2] $ and    ${\cal{O}} [ (\sigma _1 -\sigma _2 ) ^2] $. 
\section{The non local voltage}

The equation for the non local  voltage, in dimensionless units $ v_{y}=  e \: (c_{+,A}-c_{+,B}) /( k_B \:^oK) $   is derived as follows.

We go back to   Eq.(\ref{Atypeass}):
  \begin{widetext}  
   \bea
     \partial^2 _{x} (c_{+,A}-c_{+,B}) & =  g'\: (2\gamma '+g') \left [  c_{+,A}-c_{+,B} -  2 s_{+} \right ]+ g' \:  \partial _x  (s_{-,A}-s_{-,B}). \nonumber
\enea
Inserting   Eq.(\ref{eqp1}) and   Eq.(\ref{duqa})  in it,  we get ( $T \approx 1 \:^o\! K $):    
\bea
  \partial ^2_x  (c_{+,A}-c_{+,B})  - g' \: (2 \gamma '+g ') (c_{+,A}-c_{+,B})  \hspace{8cm}\nonumber\\
  = g'\left \{  - K_- \: \left ( -\partial _x^2\Delta  T \right ) -\frac{w}{4} \left (  \frac{1}{\sigma_1}- 
  \frac{1}{\sigma_2} \right )\alpha _{xy}   \partial _x\left ( \Delta T \: \frac{{\cal{E}} }{Q}
\right ) -\frac{w}{4}  \left (  \frac{1}{\sigma_1}+ \frac{1}{\sigma_2} \right )  g\:\frac{2\gamma}
{2\gamma+g }
\: \partial _x (s_{-,A}+s_{-,B}) \right \}\nonumber\\
 - g' \: (2 \gamma' +g ') \: 
 \left \{  - K_+ ( -\nabla _x \overline{T} )- \frac{w}{4} \left (  \frac{1}{\sigma_1}+ \frac{1}{\sigma_2} \right ) \: \alpha _{xy} {T} \: \frac{{\cal{E}} }{Q}
-\frac{w}{4}\left (  \frac{1}{\sigma_1}-  \frac{1}{\sigma_2} \right )
 g\:\frac{2\gamma}
{2\gamma+g }
\:  (s_{-,A}-s_{-,B})\right \} .\nonumber
 \enea
 where $ K_\pm  $ is defined in Eq.(\ref{kka}). With Eq.s(\ref{eqp1}, \ref{truq}), by  keeping   only first order in $g$, we get: 
  \bea
   \partial ^2_x  (c_{+,A}-c_{+,B})  - g' \: (2 \gamma' +g' ) (c_{+,A}-c_{+,B})  \hspace{8cm}\nonumber\\
   =  g' \: ( 2 \gamma '+g' ) \left [ K_+  \: \left ( -\partial _x\overline{ T }\right ) +\frac{w}{4}\: \frac{ \alpha _{xy}   \: 
   {\cal{E}} }{Q}  \left (  \frac{1}{\sigma_1}+ \frac{1}{\sigma_2} \right ) \: T \right ] -g'\: \partial _x\left [   K_-  \:
   \left ( -\partial _x\Delta  T \right ) +\frac{w}{4}\: \frac{ \alpha _{xy}   \: {\cal{E}} }{Q}
\left (  \frac{1}{\sigma_1}-  \frac{1}{\sigma_2} \right ) \Delta T \right ]\nonumber
 \enea
 According to  Eq.(\ref{cono}), we have chosen to impose the vanishing of the second square bracket on 
 the right hand side, so that the final equation for the Hall voltage of Eq.(\ref{vol}), reads 
  \bea
   \partial ^2_{ \tx } \:  v_{y}    - \tilde{g} \: v_{y} 
   =   \tilde{g} \:  \frac{e}{L} \:  \left (  \frac{1}{\sigma_1}+ \frac{1}{\sigma_2} \right ) \:  \frac{w}{4\:  k_B Q}\: 
   \left [ \alpha _{xy}   \: {\cal{E}} L \: \frac{T}{^o\! K} -  \left ( \kappa _{xx}  - \frac{ \sigma _1-\sigma _2}{\sigma _1+\sigma _2}\: 
   \alpha _{xy}\: Q \right ) \: \partial _{\tx} \tau  \right ]. 
   \label{Ahallvolt}
 \enea
 with $ \tilde x = x/L$, $\tau = \overline T / \:^oK $, to ${\cal{O}}[( \sigma _1-\sigma _2)^2]$.   
 \end{widetext}
\end{appendix}

%\bibliography{biblio.bib}

\begin{thebibliography}{50}
\bibitem[{not()}]{nota}
\bibinfo{note}{The definition of Eq.(\ref{imbo}) in terms of the imbalance
  chemical potential can be justified by referring to the Poisson electrostatic
  problem $ {\cal{N}} \partial ^2 \delta c /\partial y^2 = 4 \pi \rho _I $ (
  with clear meaning of the symbols), where $ \partial ^2 \delta c /\partial
  y^2 \approx \delta c / w^2 $ and $ {\cal{N}}/w^2 \propto n_\square$.}
\end{thebibliography}

\end{document}